\documentclass[preprint,aps,prd,nofootinbib]{revtex4-1}
\usepackage{graphicx} 
\usepackage{epstopdf}
\usepackage{amsmath}
\usepackage{color}
\usepackage{subcaption}
\usepackage{mathrsfs}
\usepackage{hyperref}

\usepackage{amsfonts}
\usepackage{amssymb}
\usepackage{bm}

\def\bea{\begin{eqnarray}}
\def\eea{\end{eqnarray}}
\def\<{\langle}
\def\>{\rangle}
\def\tr{\text{tr}}
\def\dtp{\delta_{\text{p}}}
\def\ordr{\mathcal{O}}
\def\A{{\text{A}}}
\def\B{{\text{B}}}
\def\prob{{\mathcal{P}}}

\def\non{\nonumber}

\def\el{{\text{el}}}
\def\inel{{\text{inel}}}
\def\tv{{\text{T}}}

\begin{document}

\title{Entanglement Entropy is Elastic Cross Section}

\author{\vspace{0.5cm} Ian Low and Zhewei Yin}
\affiliation{\vspace{0.5cm}
\mbox{High Energy Physics Division, Argonne National Laboratory, Lemont, IL 60439, USA}\\
\mbox{Department of Physics and Astronomy, Northwestern University, Evanston, IL 60208, USA}\\ 
 \vspace{0.5cm}}

\begin{abstract}

We present universal relations between entanglement entropy, which quantifies the quantum correlation between subsystems, and the elastic cross section, which is the primary observable for high energy particle scattering,  by employing a careful formulation of wave packets for the incoming particles.  For 2-to-2 elastic scattering with no initial entanglement and subdividing the system along particle labels,  we show that both the R\'enyi and Tsallis  entropies in the final states are directly proportional to the elastic cross section in  unit of the transverse size for the initial wave packets, which is then interpreted as the elastic scattering probability. The relations do not depend on the underlying dynamics of the quantum field theory and are valid to all orders in coupling strengths. Furthermore,  computing quantum correlations between momentum and non-kinematic data   leads to   entanglement entropies expressed as various semi-inclusive elastic cross sections. Our result gives rise to a novel ``area law'' for entanglement entropy in a two-body system.

\end{abstract}

\maketitle
\noindent

\tableofcontents

\section{Introduction}

Quantum entanglement is a key feature setting a quantum theory apart from classical physics \cite{Einstein:1935rr,Bell:1964kc}. Because of the intrinsic probabilistic nature of quantum physics,  there exist underlying correlations between different parts of a quantum system even when the subsystems are far apart. These quantum correlations allow one to gain knowledge of one subsystem by observing a different subsystem. Such a correlation can be quantified by the entanglement entropy \cite{Nielsen:2012yss}, which have been studied extensively in low energy, nonrelativistic systems.

At higher energies when relativistic effects become important, the theoretical framework unifying quantum mechanics and special relativity is quantum field theory (QFT). The archetypal process to consider in QFT is particle scattering in a high energy collision and the main experimental observable is the cross section, an effective area characterizing the probability of nontrivial scattering events to occur. The calculation of cross section has been a singular focus of attention in QFT since its inception. In this regard, it is somewhat surprising that entanglement entropy, as a quintessential feature of quantum mechanics and built into the foundation of QFT, has not played a prominent role in the study of particle collisions in the highly relativistic regime. In fact, experimental collaborations at the Large Hadron Collider (LHC) at CERN made the first observations of quantum entanglement only very recently in $t\bar{t}$ final states \cite{ATLAS:2023fsd,CMS:2024pts}, which constitute a verification of quantum entanglement at the highest energy to date.

The goal of the current work is to bring into sharp focus the quantumness in QFT, by studying  the  entanglement entropy in high energy 2-to-2 scattering processes, as well as  the possible relation to the single most important observable in QFT -- the cross section. It turns out that the complex and rapidly changing environment during relativistic particle scattering not only introduces complications not seen in the nonrelativistic regime but also presents calculational subtleties which need to be handled with care. Past studies in this direction mostly computed, perturbatively, the entanglement entropy in particle scatterings under specific theories \cite{Dvali:2014ila,Seki:2014cgq,Peschanski:2016hgk,Cervera-Lierta:2017tdt,Fan:2017mth,Beane:2018oxh,Tomaras:2019sjq,Dvali:2020wqi,Peschanski:2019yah,Aoude:2020mlg,Low:2021ufv,Dvali:2021ooc,Dvali:2021rlf,Muller:2022htn,Fedida:2022izl,Liu:2022grf,Cheung:2023hkq,Hentschinski:2023izh,Morales:2023gow,Carena:2023vjc,Sakurai:2023nsc,Liu:2023bnr,Hu:2024hex,Aoude:2024xpx,Kowalska:2024kbs,Colas:2024ysu},  though general, concise results are wanting.

In this work we will employ the $S$-matrix formalism and compute the entanglement entropy in a very general setting. Most importantly, we will demonstrate a simple, universal relation between  entanglement entropy and elastic cross sections:
\begin{center}
    Entanglement entropy $\sim$ Elastic cross section
\end{center}
where the meaning of $\sim$ will be made precise later. Our relation makes minimal assumptions about the specifics of the theory, and does not require any perturbative expansion in terms of the coupling strength; no  Feynman diagrams are needed in our derivation. This relation gives rise to a new ``area law''   given the dual interpretation of cross section  as an area and as a probability, as first pointed out in Ref. \cite{Low:2024mrk}.

A careful formulation of the scattering problem is needed to obtain such a relation. The standard treatment of relativistic two-body scattering in QFT assumes two incoming particles with well-defined momenta and expresses the cross section as a function of definite initial momenta. However, there are limitations of such treatment, as a momentum eigenstate in position space corresponds to a plane wave, which is infinitely spread out in the entire space. In an actual experiment, any apparatus has an experimental resolution and neither the momentum nor the position can be measured with perfect precision. Conceptually such issues are usually dealt  with by introducing wave packets \cite{Peskin:1995ev,Weinberg:1995mt}. Practically these issues are simply glossed over and computations almost always proceed with momentum eigenstates.

Understanding the subtle issues of momentum eigenstates vis-\`{a}-vis wave packets turns out to be the key to obtaining the simple relation between entanglement entropy and cross section. In this work we formulate the initial states using wave packets, which have finite characteristic sizes in both momentum and position space. The momentum eigenstate corresponds to a $\delta$-function wave packet in the momentum space and a plane wave with an infinite extent in the position space. The finite sizes of the wave packets allow us to introduce a perturbative expansion around the plane wave limit or, equivalently, the momentum eigenstate. The entanglement entropy is computed in this perturbative expansion in the size of wave packets, and the leading order results will be independent of the details of the wave packets.

What we hence discover is that in general, for 2-to-2  scatterings with no initial entanglement, the final subsystem entanglement entropy is proportional to the total elastic cross section,\footnote{Recall that the entropy for the entire system is conserved under time evolution, which is a unitary transformation.} or alternatively, is exactly the total elastic scattering probability (up to simple numerical factors). Precisely,
\begin{equation}
\text{Entanglement entropy} \quad\propto\quad \dfrac{\text{Elastic cross section}}{\text{Transverse area of wave packets}}\ , \nonumber
\end{equation}
where the proportionality constant depends on the measure of entanglement entropy, i.e. the Tsallis and R\'enyi entropies \cite{Tsallis,renyi}. The direct proportionality of the entropy to the cross section can be interpreted as an area law. Originally discovered in black hole physics \cite{Bekenstein:1973ur,Hawking:1975vcx}, area laws have been found in various quantum many-body systems \cite{Srednicki:1993im,Maldacena:1997re,Ryu:2006bv,Eisert:2008ur,Dvali:2017nis,Almheiri:2020cfm}. Our relations, however, suggest an area law for a two-body system for the first time. Furthermore, if we consider the possibility that the incoming particles can have quantum numbers other than 4-momenta, there are multiple ways to partition the system into two subsystems. It turns out that the entanglement entropy for different partitions of the two-particle system corresponds to  different semi-elastic cross sections, which will be shown later. 

The work is organized as follows. We will describe in detail our setup of the scattering problem in Sec. \ref{sec:sup}, including the description of the initial and the final states, and the formulation of wave packets and the plane wave limit. We will then derive the universal linear relations between entanglement entropy and  cross sections in Sec. \ref{sec:eetp}, implementing various possible partitions of the system. Finally, we will discuss implications of our results and further directions to explore in Sec. \ref{sec:cao}. Additional examples on initial wave packets and applications of our diagrams for kinematic information are presented in Appendices \ref{app:uwfe} and \ref{app:dkd}.

\section{The elastic scattering and the plane wave limit}

\label{sec:sup}

We will discuss the general setup of the problem in this section. In Section \ref{sec:iswp}, the wave packet formalism and definitions of entanglement entropy are introduced.  We will carefully explain  the necessity of regulating the initial momentum states using wave packets and formulate the plane wave limit. In Section \ref{sec:fses}, we compute the scattering probabilities in the wave packet formalism and demonstrate a systemaic expansion in the sizes of the wave packet to reach the plane wave limit.

\subsection{Wave packet formalism and entanglement entropies}

\label{sec:iswp}

We consider the case of scattering between two nonidentical particles labeled as type-A and type-B. Specifically the two particles differ in some intrinsic property $Q$, such as the mass, the electric charge or the spin, which renders them distinguishable from the quantum mechanical point of view. In general, the quantum number of each particle  is divided into 2 categories:
\begin{itemize}
    \item The kinematic data, labelled by the 3-momenta $\vec{p}_{\A/\B}$ and masses $m_{\A/\B}$.  The particle  can be either massive or massless. The Hilbert space representing such kinematic data is denoted by ${\cal H}_{\rm kin}$. 
    \item Non-kinematic quantum numbers $f$, which we will call ``flavors,'' whose Hilbert space is denoted ${\cal H}_{\rm f}$. Note that the ``flavors'' do not have to be internal quantum numbers, for instance the electric charge, and could be the different spin projections such as the spin-up and spin-down along a certain direction. In general, the flavor space for type-A and type-B particles can be different, and we label the flavors of type-A and -B particles using $\{ i \}$ and $\{\bar{i}\}$, respectively.
\end{itemize}
The particle label $Q$ distinguishing type-A and type-B could either be the mass or one of the flavor quantum numbers. Some concrete examples are
\begin{itemize}
\item $e^+e^-$ scattering: The natural choice for $Q$ is the electric charge which separates the electron from the positron. In this case the spin projection would be considered a ``flavor'' quantum number.
\item $e^- \mu^-$ scattering: The particle label   $Q$  in this case is the mass. 
\item $e^- \gamma$ scattering: In this case $Q$ could be either the mass or the total spin of the incoming particle.
\end{itemize}

For a system with a density matrix $\rho$, we construct a bipartite system by dividing into two subsystems, $\text{I}$ and $\text{J}$, and the reduced density matrix for subsystem I is $\rho_{\text{I}} = \tr_{\text{J}}\, \rho$. We adopt two commonly used measures of entanglement entropy: the $n$-th order Tsallis and R\'enyi entropies $\mathcal{E}_{n,\text{T}/\text{R}}$ \cite{Tsallis,renyi}, which are defined as
\bea
\mathcal{E}_{n,\text{T}} = \frac{1 - \tr\, \rho_{\text{I}}^n}{n-1},\qquad \mathcal{E}_{n,\text{R}} = \frac{1}{1-n} \log\, \tr\, \rho_{\text{I}}^n,\label{eq:etrdef}
\eea
where $n\ge 2$ is an integer. For the Tsallis entropy, the $n=2$ case is also called the linear entropy, which was employed in Ref.~\cite{Low:2024mrk}.

In this work we construct a bipartite system from 2-to-2 scattering by selecting I and J from the  set $\mbox{K}=\{p_\A, p_\B, f_\A, f_\B \}$: $\mbox{I}\in \mbox{K}$, $\mbox{J}\in \mbox{K}$ and $\mbox{I}\cup\mbox{J}= \mbox{K}$.\footnote{Further divisions, such as dividing the momentum of each particle into different magnitudes and orientations in some frame, are possible and left  for future exploration.} There are several different combinations one can choose to form $\mbox{I}$ and $\mbox{J}$: For example, one can consider $\text{I}=\{p_\A, f_\A \}$ and  $\text{J}=\{ p_\B, f_\B\}$, and in this case the entanglement entropy is a measure of the quantum correlation between type-A and type-B particles. But this is not the only possibility and different choices of $\text{I}$ and $\text{J}$ will be explored in the following section.

We consider pure initial  states, assuming the realistic case where the momenta between the two particles are not entangled, and the momenta are not entangled with the flavors either. In other words, the only entanglement we are allowing in the initial state is between $f_\A$ and $f_\B$. The state can thus be written as
\bea
\label{eq:instate}
|\text{in} \> = \sum_{i,\bar{i}} \Omega_{i \bar{i}} |\psi_{\A} \> \otimes | i\> \otimes |\psi_{\B} \> \otimes | \bar{i}\>,
\eea
where $|i \>$ and $|\bar{i} \>$ are the flavor states for type-A/B particles, respectively, which are normalized to $\<i | j \> = \delta^{ij}$ and $\<\bar{i} | \bar{j} \> = \delta^{\bar{i } \bar{j} }$; $\Omega_{i \bar{i}}$ describes the initial entanglement between the flavors of the two particles, satisfying the normalization $\tr ( \Omega^\dagger \Omega ) = 1$. On the other hand, the states $|\psi_{\A/\B}\>$ contain the kinematic data of the two initial particles. Naively, one may want to choose them as the momentum eigenstates $| p_{\A/\B}\>$, which are naturally normalized to 3-momentum $\delta$-functions:
\bea
\<p| q\> = (2 \pi )^3 2 E_{p} \delta^3 (\vec{p} - \vec{q}).\label{eq:mesn}
\eea
The initial density matrix is
\bea
\rho^{\text{i}} = | \text{in} \> \< \text{in} |,\label{eq:dmi}
\eea
which needs to be normalized such that
\bea
\tr \rho^{\text{i}} = \< \text{in} | \text{in} \> = \<\psi_\A | \psi_\A \> \<\psi_\B | \psi_\B \> = 1.
\eea
Therefore, we need $\<\psi_\A | \psi_\A \> = \<\psi_\B | \psi_\B \> = 1$, which is incompatible with the the $\delta$-function normalization in Eq. (\ref{eq:mesn}). It is thus necessary to regulate the initial state using wave packets:
\bea
\label{eq:d3pint}
| \psi \> = \int_{p} \psi (p) |p\>,\qquad \int_p \equiv \int \frac{d^3 \vec{p}}{(2 \pi)^3 \sqrt{2E_p}},
\eea
where $\psi (p)$ is the wave function, normalized as
\bea
\< \psi | \psi \> = \int \frac{d^3 \vec{p}}{(2 \pi)^3} \ |\psi (p ) |^2 = 1.\label{eq:wfnrq}
\eea

The wave function $\psi (p)$ can be completely general; however, we are  interested in the plane wave limit, i.e. when the wave functions of the initial particles are sharply peaked around some momentum $k_{\A/\B}$, which is usually how a scattering experiment is set up. In this case, it is possible to define a center-of-mass (CoM) frame where $\vec{k}_\A + \vec{k}_\B = 0$, and it is convenient to choose $\vec{k}_\A = - \vec{k}_\B = \vec{k}$ in the $p_z$-direction, i.e. $\vec{k} = (0,0,|\vec{k}|)$. We can then define a 3-dimensional ``peak'' function $\tilde{\delta}^3 (\vec{p} )$ in the momentum space with characteristic widths of $\dtp$ in the longitudinal direction and $\delta_\tv$ in the transverse direction. Notice that $L \equiv 1/\delta_\tv$ is the transverse, linear size of the position space wave packet. In relativistic scatterings we expect 
 \bea
 \frac1{L}=\delta_\tv \ll \dtp\ , 
 \eea
 due to the following observation: The wave functions in the position space will have widths of $\ordr(1/\dtp)$ in the longitudinal and $\ordr (1/\delta_\tv) = \ordr (L)$ in the transverse directions, and during relativistic scatterings the wave functions are Lorentz contracted in the direction of motion.\footnote{In the position space the wave packet can be considered a pancake-like object transverse to the direction of motion.} Thus it is natural to expect $1/\dtp \ll 1/\delta_\tv=L$ for a relativistic particle. As such, the plane wave limit is reached via
 \bea
 \lim_{\dtp \to 0 } \tilde{\delta}^3 (\vec{p} ) = \delta^3 (\vec{p}).\label{eq:pfdef}
\eea
 Furthermore, from the normalization of the $\delta$-function in Eq. (\ref{eq:pfdef}), we can estimate the height of the peak function as
\bea
\tilde{\delta}^{3} (\vec{0}) \sim \frac{1}{\dtp \delta_\tv^2}.
\eea
The properly normalized initial wave functions are then
\bea
\psi_{\A/\B} (\vec{p}) = \mathcal{N} e^{-i \vec{p} \cdot \vec{x}_{\A/\B}} \  \tilde{\delta}^3 ( \vec{p} - \vec{k}_{\A/\B} ),\label{eq:gwpn}
\eea
where $\mathcal{N} $ is an $\ordr (\delta_\tv\sqrt{\dtp})$ constant, while $\vec{x}_{\A/\B}$ gives the  transverse displacement of the two incoming particles. We can choose 
\begin{equation}
    \vec{x}_{\B} = - \vec{x}_{\A} = \vec{b}/2 \ ,
\end{equation}
where $\vec{b}$ is the impact parameter and can be chosen such that $b_z = 0$. The head-on collision corresponds to $\vec{b}=0$.

\begin{figure}[tbp]
     \centering
     \includegraphics[width=0.3\textwidth]{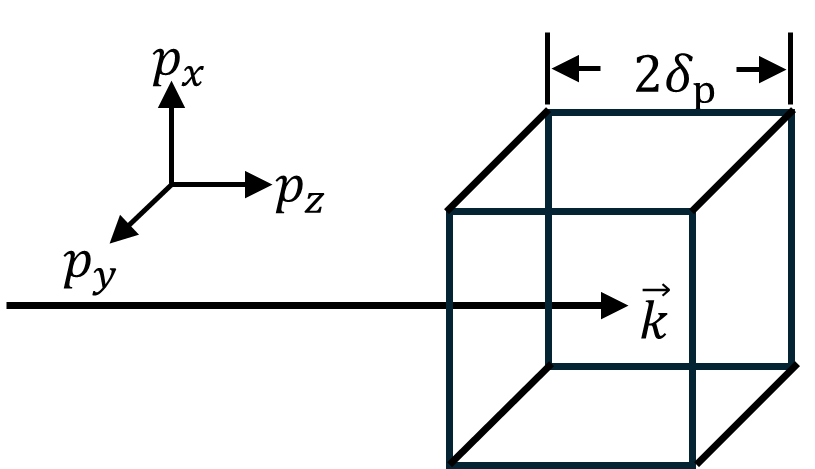}
     \caption{The form of wave packets we choose in our example, shown in   momentum space. The wave function is strictly confined inside a cube of side length $2 \dtp$.\label{fig:wpms}}
     \end{figure}

We will argue that the leading order calculation of the entanglement entropy in the plane wave limit is independent of the details of  the peak function $\tilde{\delta}^3$, and  provide reasonings that work for general wave functions in the plane wave limit. As an illustration  we  provide an explicit example of initial wave packets, chosen to be bounded inside a cube of side length $2 \dtp$ in  momentum space, as shown in Fig. \ref{fig:wpms}. More explicitly,
\bea
\tilde{\delta}^3 (\vec{p}) &=& \tilde{\delta}_0 (p_z) \tilde{\delta}^2_\tv (\vec{p}_\tv), \label{eq:deltil2} \\
\tilde{\delta}_{0} (l) &\equiv&   \frac{\Theta ( l + \dtp) -  \Theta ( l - \dtp) }{2 \dtp},\label{eq:deltil}\\
 \tilde{\delta}^2_\tv (\vec{p}_\tv) &\equiv& \frac{1}{\mathcal{N}_\tv} \left[  \Theta ( p_x + \dtp) -  \Theta ( p_x - \dtp)  \right] \left[  \Theta ( p_y + \dtp) -  \Theta ( p_y - \dtp) \right] \int_{A_{\vec{r}}} d^2 \vec{r}_\tv \, e^{-i\vec{p}_\tv \cdot \vec{r}_\tv},\ \ \label{eq:guxpd}
\eea
where $\vec{p}_\tv = (p_x, p_y, 0)$ is the transverse component of $\vec{p}$ and $\int_{A_{\vec{r}}}$ denotes an integral over a transverse area $A_{\vec{r}}$ centered at $\vec{r}$, the location of the particle. In addition,
\bea
\mathcal{N}_\tv = \int_{-\dtp}^{\dtp} dp_x \int_{-\dtp}^{\dtp} dp_y \int_{A_{\vec{r}}} d^2 \vec{r}_\tv \, e^{-i\vec{p}_\tv \cdot \vec{r}_\tv}\label{eq:ntdux}
\eea
is a normalization factor such that
\bea
\int d^2 \vec{p}_\tv \,\tilde{\delta}^2_\tv (\vec{p}_\tv) = 1.
\eea
Observe that the wave function in the longitudinal direction, Eq. (\ref{eq:deltil}),  is simply a ``box function'' consisted of two step functions. In the transverse direction, Eq. (\ref{eq:guxpd}) is  two box functions multiplied by the ``partial'' Fourier transform of a ``constant'' function in position space, in that we are only integrating over a finite size area $A_{\vec{r}}$. Therefore, the position wave packet in the transverse direction, given by the inverse Fourier transform of Eq. (\ref{eq:guxpd}), is  approximately (but not exactly) uniform inside $A_{\vec{r}}$, due to the finite size of the box in momentum space. In Ref. \cite{Low:2024mrk} we  chose $A_{\vec{r}}$  to be a square  and $L^2$ is precisely the area of $A_{\vec{r}}$.  In this work we will keep the shape of $A_{\vec{r}}$ general whenever possible, in which case $L^2=1/\delta_\tv^2$ is still an ``effective measure'' of the area of $A_{\vec{r}}$.  A different example of momentum wave packet, which consists of three simple box functions in all three directions,  is presented in Appendix \ref{app:uwfe}. 

\begin{figure}[tbp]
 \centering
    \subfloat[\centering \label{fig:wpdz}]{{\includegraphics[width=0.46\textwidth]{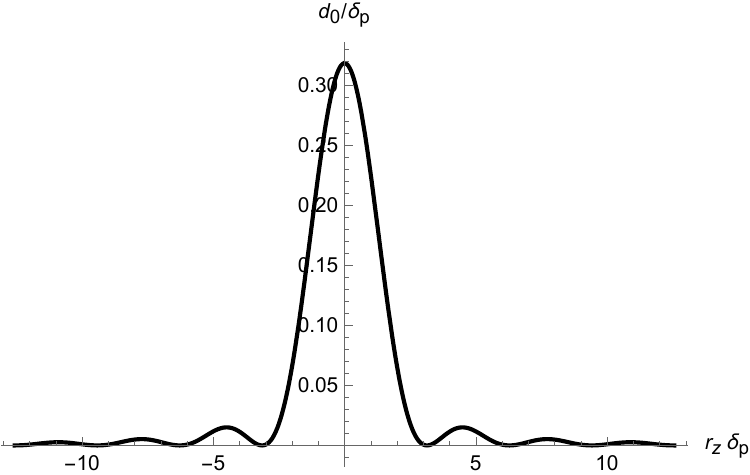} }}%
    \qquad
    \subfloat[\centering \label{fig:wpdt}]{{\includegraphics[width=0.46\textwidth]{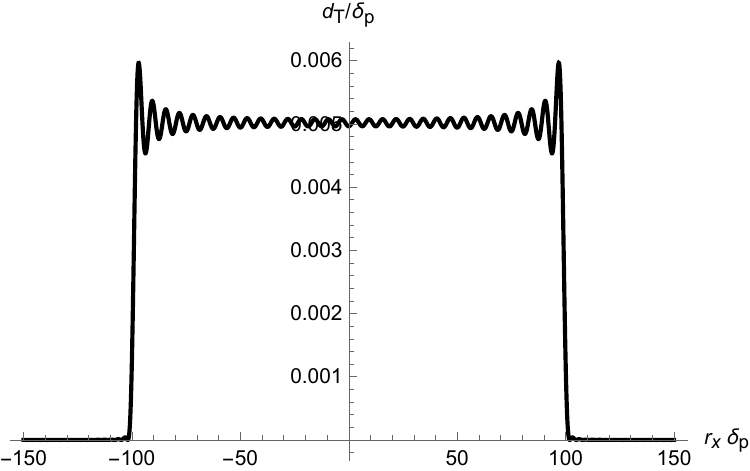} }}%
    \caption{(a) The longitudinal field density $d_0(r_z)$ in position space. (b) The transverse field density $d_\tv (r_x)$ in the $x$-direction  for $A_{\vec{x}_{\A}} $ being a square bounded by $|r_x|,|r_y| \le L/2$, with $L\dtp =200$.}
\end{figure}

Given our choice of wave packets, the normalization factor in Eq. (\ref{eq:gwpn}) is now ${\cal N}= \sqrt{{4\pi \dtp}/{\mathcal{N}_\psi}}$, where
\bea
\mathcal{N}_\psi = \int_{-\dtp}^{\dtp} dp_x \int_{-\dtp}^{\dtp} dp_y \frac{1}{\mathcal{N}_\tv^2 (2\pi)^2} \left( \int_{A_{\vec{0}}} d^2 \vec{r}_\tv \, e^{-i\vec{p}_\tv \cdot \vec{r}_\tv} \right)^2 = \frac{L^2}{(2 \pi)^4} \left[1 +  \ordr ( L^{-1}/\dtp)\right] .\label{eq:npslim}
\eea
In the last equality of the above, we have utilized the condition $1/L \ll \dtp$, so that the bounds of the momentum integrals in Eqs. (\ref{eq:ntdux}) and (\ref{eq:npslim}) can be extended to infinity. It would be illuminating to plot the field density in position space for the simple case of $A_{\vec{r}}$ being a square of side length $L$, by computing the wave function $\tilde{\psi}_\A (\vec{r})$ in position space, which is the Fourier transform of $\psi_{\A} (\vec{p})$. Choosing $\vec{b}=0$ and $A_{\vec{x}_{\A}=\vec{0}}$  to be a square bounded by $|r_x|, |r_y| \le L/2$, the field density factorizes:
\bea
\left| \tilde{\psi}_\A (\vec{r}) \right|^2 = d_{\text{T}} (r_x) d_{\text{T}} (r_y) d_0 (r_z).
\eea
We plot the longitudinal density $d_0 (r_z)$  and transverse density $d_{\text{T}} (r_x)$ in the $x$-direction in Figs. \ref{fig:wpdz} and \ref{fig:wpdt}, respectively.  As the momentum space wave function in the $p_z$-direction is uniform within a $2\dtp$ sized window, the field density in position space is an unbounded peak of $\ordr (1/\dtp)$ width, as shown in Fig. \ref{fig:wpdz}. On the other hand, Fig. \ref{fig:wpdt} shows how in the $x$-direction the position space field density is roughly uniform inside $|r_x| \le L/2$ and roughly $0$ outside, where we have chosen $1/L=\dtp/200$.

After these discussions, it should be clear now that the plane wave limit is given by
\begin{equation}
\label{eq:planelimit}
    \frac{\dtp}{|\vec{k}|} \longrightarrow 0\ , \qquad
    \frac1{L\,\dtp} \ll 1 \ , \qquad  \frac1{L\,\dtp} \  \lesssim\  \frac{\dtp}{|\vec{k}|}\ ,
\end{equation}
where the last relation is a parametric choice such that there is a single small parameter to expand.

\subsection{The final state for elastic scattering}

\label{sec:fses}

In general, the scattering process is described by the $S$-matrix: $S = 1 + iT$, where the transition matrix $T$ is related to the scattering amplitude $M$ through
\bea
 \< \{k_{\text{f}} \}, f_{\text{f}}| T | \{k_{\text{i}}\}, f_{\text{i}} \>= (2 \pi)^4 \delta^{4} \left(\sum k_{\text{f}} - \sum k_{\text{i}} \right) M_{f_{\text{i}}, f_{\text{f}}} (\{k_{\text{i}}\}; \{k_{\text{f}} \}),\label{eq:rtm}
\eea
with $\{ k_{\text{i}/\text{f}} \}$ and $f_{\text{i}/\text{f}}$ describing the momentum and flavor configurations of the initial and final states. Unitarity of the $S$-matrix, $S^\dagger S = 1$, leads to the optical theorem, $2\, \text{Im}\, T = T^\dagger T$,  which will be heavily utilized in the following.

The result of the scattering  is described by an out-state given by  $|\text{out} \> = S|\text{in} \>$, which in general is a linear superposition of all possible outcomes, including one-particle, two-particle, and multi-particle final states allowed by the kinematics and the underlying interactions. We will be focusing on  the \textit{elastic} scattering,
\begin{equation}
    {\rm A} + {\rm B} \longrightarrow   {\rm A} + {\rm B} \ ,
\end{equation}
where the final state consists of exactly one type-A particle and one type-B particle. This is only a subset of all possible outcomes. All other processes constitute what we call  \textit{inelastic} scattering.  Let us illustrate with  explicit examples.  In $e^+(\uparrow) e^-(\downarrow)$ scattering, we choose particle A to be the positron, particle B to be the electron, and the ``flavor'' being the spin orientation. Then the elastic processes include e.g. $e^+(\uparrow) e^-(\downarrow) \to e^+(\uparrow) e^-(\downarrow)$, which preserves the flavor for both of the particles; $e^+(\uparrow) e^-(\downarrow) \to e^+(\downarrow) e^-(\uparrow)$, which changes the flavors of both of the particles; and $e^+(\uparrow) e^-(\downarrow) \to e^+(\downarrow) e^-(\downarrow)$, which only changes the flavor of particle A. Examples for inelastic processes include $e^+e^- \to \mu^+ \mu^-$, as the final particles do not have the same mass as the initial particles, as well as $e^+ e^- \to e^+ e^- \gamma$, which has an extra particle in the final state.

In the following we introduce a projection operator $P_{\A \B}$, which
selects the part of the Hilbert space where each state contains exactly one type-A and one type-B particles. The final state that we are interested in is then $|\text{out}\>_{\el} = P_{\A \B} | \text{out}\>$. Although the full final state is properly normalized to 1 because of the unitarity of the $S$-matrix: $\< \text{out} | \text{out} \> = \< \text{in} | S^\dagger S| \text{in} \> = 1$, $P_{\A \B} $ corresponds to a measurement which collapses the wave function. Following the L\"uders rule \cite{luders1950zustandsanderung}, the properly normalized density matrix after applying the projection $P_{\A \B}$ is
\bea
\rho^{\text{f}} = \frac{P_{\A \B} |\text{out} \> \< \text{out}| P_{\A \B}}{ \tr\, (P_{\A \B} |\text{out} \> \< \text{out}|)} = \frac{1}{1-\prob_{\inel}}|\text{out} \>_{\rm el}\ _{\rm el} \< \text{out}|\ , \label{eq:fdmpn}
\eea
where
\bea
\prob_{\inel} = \< \text{out} |1-P_{\A \B} | \text{out} \>= \< \text{in}| T^\dagger (1 - P_{\A\B}) T |\text{in} \>\label{eq:c2d}
\eea
is the \textit{probability} of inelastic scattering. The probability $\prob_{\el}$ for elastic scattering and the probability $\prob_{\text{tot}}$ for any scattering to occur are similarly given by
\bea
\label{eq:peltot}
\prob_{\el} =  \< \text{in}| T^\dagger  P_{\A\B} T |\text{in} \>, \qquad \prob_{\text{tot}} = \< \text{in}| T^\dagger   T |\text{in} \>.
\eea
Clearly $\prob_{\el} + \prob_{\inel} = \prob_{\text{tot}}< 1$, as $1- \prob_{\text{tot}}$ describes the probability that no scattering happens. Next we compute the probabilities in the plane wave limit in Eqs. (\ref{eq:planelimit}), by expanding around the small  dimensionless parameter $\dtp/|\vec{k}|$ and  keeping the leading non-vanishing results. 

We use $\prob_{\text{tot}}$ in Eq. (\ref{eq:peltot}) as an illustration. By plugging in the in-state in  Eq. (\ref{eq:instate}) and keeping the general form of the wave packets for now, $\prob_{\text{tot}}$  becomes,
\bea
\prob_{\text{tot}} =  \int_{\substack{p_1, p_2, \\q_1, q_2}}   \psi_\A (p_1) \psi_\B (p_2) \psi_\A^* (q_1)   \psi_\B^* (q_2) (2\pi)^4 \delta^4 (q_1 + q_2 - p_1 - p_2)\, {\cal F}_{\text{tot}} (p_1, p_2, q_1, q_2)\, ,\label{eq:ptotitg}
\eea
where
\bea
\label{eq:ftotome}
(2\pi)^4 \delta^4 (q_1 + q_2 - p_1 - p_2)\, {\cal F}_{\text{tot}} (p_1, p_2, q_1, q_2)
 \equiv \sum_{i,\bar{i},j,\bar{j}} \Omega_{i \bar{i}} \Omega^*_{j \bar{j}}\<q_1,j;q_2,\bar{j}|T^\dagger T|p_1,i;p_2,\bar{i}\>\, .
\eea
We have inserted a momentum-conserving delta function on the left-hand side of the above, anticipating that ${\cal F}_{\text{tot}}$ can be expressed in terms of amplitudes using Eq. (\ref{eq:rtm}). Keeping in mind that the incoming particles have momentum wave functions centered at $\vec{k}_{\A}$  and $\vec{k}_{\B}$, we Taylor-expand ${\cal F}_{\text{tot}}$ around $k_{\A/\B}$ as
\bea
{\cal F}_{\text{tot}} (p_1, p_2, q_1, q_2) &=& {\cal F}_{\text{tot}} (k_\A, k_\B, k_\A, k_\B) + \sum_{n_1,n_2,n_3,n_4 = 1}^{+\infty} \frac{1}{n_1 ! n_2 ! n_3 ! n_4 !} [ (\vec{p}_1 - \vec{k}_\A) \cdot \vec{\nabla}_{p_1} ]^{n_1} \non\\
&&\times [ (\vec{p}_2 - \vec{k}_\B) \cdot \vec{\nabla}_{p_2} ]^{n_2}[ (\vec{q}_1 - \vec{k}_\A) \cdot \vec{\nabla}_{q_1} ]^{n_3} [ (\vec{q}_2 - \vec{k}_\B) \cdot \vec{\nabla}_{q_2} ]^{n_4} \non\\
&&\times {\cal F}_{\text{tot}} (p_1, p_2, q_1, q_2)\Big|_{p_1 , q_1 = k_\A; p_2, q_2 = k_\B} \ ,\label{eq:tee}
\eea
where ${\cal F}_{\text{tot}} (k_\A, k_\B, k_\A, k_\B)$ is independent of $\{p_1,p_2, q_1,q_2\}$ and can be pulled outside of the wave function integration in Eq. (\ref{eq:ptotitg}). The subleading terms in Eq. (\ref{eq:tee}), when convoluted with the momentum wave packets in Eq. (\ref{eq:ptotitg}), give rise to integrals involving higher moments of the wave functions because of factors like $[ (\vec{p}_1 - \vec{k}_\A) \cdot \vec{\nabla}_{p_1} ]^{n_1}$ in the Taylor expansion. These higher moments are  highly suppressed in the plane wave limit if the momentum wave packet is exponentially suppressed, or vanishes identically, at large $|\vec{p}-\vec{k}_{\A/\B}|$. In this case their contributions are only $\ordr (\dtp^{n_1+n_2+n_3+n_4})$, compared to the leading term, and can be safely neglected. This is the reason we choose to strictly confine the peak functions within the box given by Fig. \ref{fig:wpms} in our examples.

In the end, keeping only ${\cal F}_{\text{tot}} (k_\A, k_\B, k_\A, k_\B)$ in Eq. (\ref{eq:tee}), 
\bea
{\prob}_{\text{tot}} &=&  \frac1{4 |\vec{k} | \sqrt{s}}\, {\cal F}_{\text{tot}} (k_\A, k_\B, k_\A, k_\B)\, I_0 (|\vec{k}|)\, \left[1 + \mathcal{O} (\dtp/|\vec{k}|) \right]\ , 
\label{eq:sptot}
\eea
where
\bea
I_0 (|\vec{k}|) \equiv 4 |\vec{k} | \sqrt{s} \int_{\substack{p_1, p_2, \\q_1, q_2}}   \psi_\A (p_1) \psi_\B (p_2) \psi_\A^* (q_1)   \psi_\B^* (q_2) (2\pi)^4 \delta^4 (q_1 + q_2 - p_1 - p_2) \label{eq:i0def}
\eea
is a universal integral of wave functions that, as we will see, appears in all of our leading order results in the plane wave limit. Moreover, inserting a complete basis of states into the righ-hand side of Eq. (\ref{eq:ftotome}), it is straightforward to show that 
\bea
{\cal F}_{\text{tot}} (k_\A, k_\B, k_\A, k_\B)& =& \sum_\text{f} \int d\Pi_\text{f}  \ (2\pi)^4 \delta^4 \left(k_\A + k_\B - \sum p_f\right) \Big| \sum_{i,\bar{i}} \Omega_{i \bar{i}} M_{i\bar{i}, f_{\text{f}}} (k_\A,k_\B;\{p_\text{f} \}) \Big|^2\non\\
& =& 4 |\vec{k} | \sqrt{s}\, \sigma_{\text{tot}}\, ,
\eea
where $\sum_{\text{f}}$  in the first line is over all possible final state $\text{f}$ and $\sigma_{\text{tot}}$ is the total scattering cross section.\footnote{This is part of the derivation of optical theorem in QFT!} In the end, we obtain a  simple relation in the plane wave limit:
\begin{equation}
    {\cal P}_{\rm tot} = I_0 (|\vec{k}|)  \left[ \sigma_{\text{tot}} +\ordr (\dtp/|\vec{k}|) \right]\,   .\label{eq:sptot1}
\end{equation}
Similarly, we have
\bea
{\prob}_{\el} = I_0 (|\vec{k}|)  \left[ \sigma_{\el} +\ordr (\dtp/|\vec{k}|) \right],\quad {\prob}_{\inel} = I_0 (|\vec{k}|)  \left[ \sigma_{\inel} +\ordr (\dtp/|\vec{k}|) \right],\label{eq:eliegp}
\eea
where $\sigma_{\el/\inel}$ is the total elastic/inelastic cross section, given by
\bea
\sigma_\el &=& \frac{(2\pi)^4}{4 |\vec{k} | \sqrt{s}} \sum_{\text{f} \in \{\A\B\}} \int d\Pi_\text{f}  \  
\delta^4 \left(k_\A + k_\B - \sum p_f\right) \Big| \sum_{i,\bar{i}} \Omega_{i \bar{i}} M_{i\bar{i}, f_{\text{f}}} (k_\A,k_\B;\{p_\text{f} \}) \Big|^2,\\
\sigma_\inel &=& \frac{(2\pi)^4 }{4 |\vec{k} | \sqrt{s}} \sum_{\text{f} \notin \{\A\B\}} \int d\Pi_\text{f}  \ \delta^4 \left(k_\A + k_\B - \sum p_f\right) \Big| \sum_{i,\bar{i}} \Omega_{i \bar{i}} M_{i\bar{i}, f_{\text{f}}} (k_\A,k_\B;\{p_\text{f} \}) \Big|^2\, .\ 
\eea
In the above $\{\A \B\}$ are the collection of states that consist of exactly one type-A and one type-B particle. Obviously, $\sigma_{\text{tot}} = \sigma_\el + \sigma_{\inel}$.

We expect the relations of $\prob \approx I_0 (|\vec{k}|) \,\sigma$ in Eqs. (\ref{eq:sptot1}) and (\ref{eq:eliegp}) to be  \textit{universal}, as long as we consider the leading order of the plane wave limit, with the higher moments in Eq. (\ref{eq:tee}) sufficiently suppressed. This means that the relations do not depend on details of the wave packet, which are all included in the overall factor $I_0 (|\vec{k}|)$. They can be viewed as a factorization between the 
input of the initial conditions and the underlying dynamics of the process.

The next step is to  evaluate $I_0 (|\vec{k}|)$ in the plane wave limit. We first provide a general argument for the scaling behavior of $I_0 (|\vec{k}|)$, and then compute it explicitly using our choice of wave packets.  Recalling the integration measure defined in Eq. (\ref{eq:d3pint}),  $I_0 (|\vec{k}|)$  in Eq. (\ref{eq:i0def}) is, schematically,
\bea
\label{eq:i0schem}
I_0 (|\vec{k}|) \sim \Big( \int dp_\tv \Big)^8 \Big( \int dp_z \Big)^4 \psi^4 \, \delta^4 \Big(\sum p \Big) \sim \Big( \int dp_\tv \Big)^6 \Big( \int dp_z \Big)^2 \psi^4\ ,\label{eq:i0gsitm}
\eea
where $\int d p_{\rm T}$ denotes a 1-dimensional integration in the transverse direction of some momentum variable. In the high energy limit the energy of each particle, $E_{\A/\B}$, is approximately the same as  $|\vec{p}|$. Since $\vec{p}$  lies predominantly in the $p_z$-direction,  we  treat the energy-conserving $\delta$-function in the above as an additional $\delta$-function in the $p_z$-direction, which  removes one additional $p_z$ integration.

Since the wave packets are narrowly concentrated around the incoming momenta of the particles,
the integrand in  Eq. (\ref{eq:i0schem}) has support only in the window of $\ordr (\dtp)$ in the $p_z$-direction and $\ordr (\delta_\tv)$ in the transverse directions, which motivates the following scaling:
\bea
\int dp_\tv \sim \delta_\tv = \frac1L\ , \qquad \int dp_z \sim \dtp\ .\label{eq:sb1di}
\eea
On dimensional ground, each wave packet in the integrand contributes $\ordr (1/\sqrt{\delta_\tv^2 \dtp})$, as seen in Eq. (\ref{eq:gwpn}). Then Eq. (\ref{eq:i0gsitm}) becomes
\bea
I_0 (|\vec{k}|) \sim \delta_\tv^6 \dtp^2 \frac{1}{(\delta_\tv^2 \dtp)^2} \sim \delta_\tv^2 = \frac1{L^2}\ ,\label{eq:i0gsc}
\eea
Therefore, $I_0 (|\vec{k}|)$ is the inverse of the characteristic transverse area of the {\em position} wave function. Notice that at the leading order $I_0 (|\vec{k}|)$ actually is independent of the width $\dtp$ in the $p_z$-direction, which is a part of a general pattern that we will also see in the following section.

From the scaling argument we immediately see from Eqs. (\ref{eq:sptot}) and (\ref{eq:eliegp}) that all  scattering probabilities are suppressed by the transverse size of the wave packet in the position space in the plane wave limit. This can be understood from the observation that   we are scattering only two particles and, in the strict plane wave limit when $L\to \infty$, the probability for the two particles to actually collide with each other becomes vanishingly small; in this limit they just fly pass each other. This also suggests that, in Eq. (\ref{eq:fdmpn}), the inelastic probability $\prob_\inel$ in the denominator of $\rho^{\text{f}}$ can be treated as a small quantity in the plane wave limit,
\begin{equation}
    \frac1{1-\prob_\inel} \approx 1 + \prob_\inel\ ,
\end{equation}
which we employ when computing the entanglement entropy.

Now we compute $I_0 (|\vec{k}|)$ explicitly.  Plugging in Eq. (\ref{eq:gwpn}) to Eq. (\ref{eq:i0def}) we have
\bea
I_0 (|\vec{k}|) &=& 4 |\vec{k}| \sqrt{s} \int d^3 \vec{p}_1 d^3 \vec{p}_2 d^3 \vec{q}_1 d^3 \vec{q}_2 \   {\cal F}_0 (p_1, p_2, q_1, q_2) e^{-i (\vec{p}_2 - \vec{p}_1 - \vec{q}_2 + \vec{q}_1) \cdot \vec{b}/2} \non\\
&&\times \delta^{4} (q_1 + q_2 - p_1 - p_2) \tilde{\delta}^3 (\vec{p}_1 - \vec{k}) \tilde{\delta}^3 (\vec{p}_2 + \vec{k}) \tilde{\delta}^3 (\vec{q}_1 - \vec{k}) \tilde{\delta}^3 (\vec{q}_2 + \vec{k}) ,\label{eq:i0ci}
\eea
where
\bea
{\cal F}_0 (p_1, p_2, q_1, q_2) = \frac{|\mathcal{N}|^4}{(2 \pi)^8 4\sqrt{E_{p_1} E_{p_2} E_{q_1} E_{q_2}}}.
\eea
Since the wave packet only has support  inside small cubes of side lengths $2 \dtp$ in momentum space,  we can write
\bea
{\cal F}_0 (p_1, p_2, q_1, q_2) = \frac{|\mathcal{N}|^4}{(2 \pi)^8 2 E_{k_\A} 2 E_{k_{B}}} [1 + \mathcal{O} (\dtp/|\vec{k}|) ]\ 
\eea
and pull ${\cal F}_0$ out of the integral in Eq. (\ref{eq:i0ci}):
\bea
I_0 (|\vec{k}|) &=&    \frac{|\vec{k}|\, \sqrt{s}\,|\mathcal{N}|^4}{(2 \pi)^8  E_{k_\A}  E_{k_{B}}} [1 + \mathcal{O} (\dtp/|\vec{k}|) ] \int d^3 \vec{p}_1 d^3 \vec{p}_2 d^3 \vec{q}_1 d^3 \vec{q}_2 \    e^{-i (\vec{p}_2 - \vec{p}_1 - \vec{q}_2 + \vec{q}_1) \cdot \vec{b}/2} \non\\
&&\times\delta^{4} (q_1 + q_2 - p_1 - p_2) \tilde{\delta}^3 (\vec{p}_1 - \vec{k}) \tilde{\delta}^3 (\vec{p}_2 + \vec{k}) \tilde{\delta}^3 (\vec{q}_1 - \vec{k}) \tilde{\delta}^3 (\vec{q}_2 + \vec{k}).
\eea
Then we remove $d^3\vec{p}_1$ using the 3-momentum $\delta$-function, and  $d(p_2)_z$ using the energy $\delta$-function:
\bea
\label{eq:energydel}
\delta (E_{p_1} + E_{p_2} - E_{p_3} - E_{p_4})  \to \left( \frac{E_{k_\A} E_{k_\B}}{ |\vec{k}| \sqrt{s} } + \mathcal{O} (\dtp/|\vec{k}|) \right) \delta ( (p_2)_z - r_0),
\eea
where   $r_0$ is the root of $(p_2)_z$ in the following equation:
\bea
\sqrt{(\vec{q_1} + \vec{q_2} - \vec{p_2})^2 + m_\A^2 } + \sqrt{\vec{p}_2^2 + m_\B^2} - \sqrt{\vec{q}_1^2 + m_\A^2} - \sqrt{\vec{q}_2^2 + m_\B^2} = 0\, .\label{eq:edfrz}
\eea
Solving  $r_0$ to the first order  in $\dtp$, we have
\begin{equation}
    r_0 = (q_2)_z + \mathcal{O} (\dtp^2/|\vec{k}|^2)\ ,
\end{equation}
which effectively turns the energy-conserving delta function $ \delta ( (p_2)_z - r_0)$ in Eq. (\ref{eq:energydel}) into $\delta [(\vec{p}_2 - \vec{q}_2)_z]$. Moreover, the coefficient of $\delta ( (p_2)_z - r_0)$ on the right-hand side of Eq. (\ref{eq:energydel}) can again be pulled out of the integral to the leading order in the plane wave limit. In the end the $p_z$-integration  decouples from the transverse directions, and plugging in $\tilde{\delta}_{0} (l)$ in Eq. (\ref{eq:deltil}), we arrive at
\bea
\label{eq:pzint}
&& \int d (p_1)_z\, d (p_2)_z\, d (q_1)_z\, d (q_2)_z \ \tilde{\delta}_0 (\vec{p}_1 - \vec{k}) \,\tilde{\delta}_0 (\vec{p}_2 + \vec{k}) \,\tilde{\delta}_0 (\vec{q}_1 - \vec{k}) \,\tilde{\delta}_0 (\vec{q}_2 + \vec{k})\non \\
&& \times \delta \left[( \vec{p}_1+ \vec{p}_2- \vec{q}_1 - \vec{q}_2)_z \right]\ \delta (E_{p_1} + E_{p_2} - E_{p_3} - E_{p_4}) \non \\
=&&\frac{1}{(2\dtp)^4}\int_{|\vec{k}|-\dtp}^{|\vec{k}|+\dtp} d (p_1)_z \int_{-|\vec{k}|-\dtp}^{-|\vec{k}|+\dtp} d (p_2)_z  \int_{|\vec{k}|-\dtp}^{|\vec{k}|+\dtp} d (q_1)_z \int_{-|\vec{k}|-\dtp}^{-|\vec{k}|+\dtp} d (q_2)_z  \non\\
 &&\times  \delta \left[( \vec{p}_1+ \vec{p}_2- \vec{q}_1 - \vec{q}_2)_z \right] \delta [(\vec{p}_2 - \vec{q}_2)_z]  \left( \frac{E_{k_\A} E_{k_\B}}{ |\vec{k}| \sqrt{s} } + \mathcal{O} (\dtp/|\vec{k}|) \right)\nonumber\\
 =&& \left( \frac{E_{k_\A} E_{k_\B}}{ |\vec{k}| \sqrt{s} } + \mathcal{O} (\dtp/|\vec{k}|) \right) (2 \dtp)^{-2} ,
\eea
where we have used Eq. (\ref{eq:energydel}) in the first equality.

For the transverse directions, we need to evaluate
\bea
&&\int d^2 \vec{p}_{1 \tv}\, d^2 \vec{p}_{2\tv}\, d^2 \vec{q}_{1 \tv}\, d^2 \vec{q}_{2 \tv} \    e^{-i (\vec{p}_2 - \vec{p}_1 - \vec{q}_2 + \vec{q}_1)_\tv \cdot \vec{b}/2}\ \delta^{2} \left[ (\vec{q}_1 + \vec{q}_2 - \vec{p}_1 - \vec{p}_2)_\tv \right] \non\\
&&\times\ \tilde{\delta}^2_\tv \left[ (\vec{p}_1 - \vec{k})_\tv \right]\ \tilde{\delta}^2_\tv \left[ (\vec{p}_2 + \vec{k})_\tv \right]\ \tilde{\delta}^2_\tv \left[ (\vec{q}_1 - \vec{k})_\tv \right]\ \tilde{\delta}^2_\tv \left[ (\vec{q}_2 + \vec{k})_\tv \right],\label{eq:c2tio}
\eea
where the peak functions are given in Eq. (\ref{eq:guxpd}). In the plane wave limit given by Eq. (\ref{eq:planelimit}), we can take advantage of $1/L \ll \dtp$ when performing the momentum integrals, similar to what we did in Eq. (\ref{eq:npslim}), so that Eq. (\ref{eq:c2tio}) becomes
\bea
&&[1 + \mathcal{O} (\dtp/|\vec{k}|) ] \frac{1}{(2 \pi)^2} \int_{A_{\vec{0}}} d^2 \vec{x}_{1 \tv} \int_{A_{\vec{0}}} d^2 \vec{x}_{2 \tv} \  \delta^2  \left[ (\vec{x}_1 - \vec{x}_2)_\tv - \vec{b}\right]\non\\
& =&\frac{L^2}{(2 \pi)^2} \left[ \mathcal{R} ( \vec{b}) + \mathcal{O} (\dtp/|\vec{k}|) \right]  .
\eea
In the above
\bea
\mathcal{R} ( \vec{b}) \equiv \frac{1}{L^2} \int_{A_{\vec{b}}} d^2 \vec{x}_{1 \tv} \int_{A_{\vec{0}}} d^2 \vec{x}_{2 \tv} \delta^2  \left[ (\vec{x}_1 - \vec{x}_2)_\tv \right]\ ,
\eea
which  characterizes the overlap of the two incoming wave functions in the transverse directions. Recall the impact parameter $\vec{b}= \vec{x}_\B - \vec{x}_\A$. For head-on collisions $\vec{b}=\vec{0}$ and $\mathcal{R} ( \vec{0}) = 1$,  while for large enough impact parameters (i.e., $|\vec{b}| \gtrsim {\cal O}(L)$)  the initial wave functions of the two particle do not overlap at all and ${\cal R}\to 0$. Including the normalization factor, $\mathcal{N} =  \sqrt{4 \pi \dtp /\mathcal{N}_\psi}$, we have
\bea
I_0 (|\vec{k}|) &=&\frac{|\vec{k}| \sqrt{s}|\mathcal{N}|^4}{(2 \pi)^8  E_{k_\A}  E_{k_{B}}} \frac{E_{k_\A} E_{k_\B}}{ |\vec{k}| \sqrt{s} } \frac{|A|}{(2 \dtp)^2 (2 \pi)^2} \left[ \mathcal{R} ( \vec{b}) + \mathcal{O} (\dtp/|\vec{k}|) \right]\non\\
&  = & \frac{1}{|A|} \left[\mathcal{R} ( \vec{b} )+ \mathcal{O} (\dtp/|\vec{k}|) \right].\label{eq:ic2g}
\eea
For $\mathcal{R} ( \vec{b} ) \sim \ordr (1)$, the above scales as $\ordr (1/|A|) = \ordr (1/L^2) = \ordr (\delta_\tv^2)$,  confirming Eq. (\ref{eq:i0gsc}). 

Given Eq. (\ref{eq:ic2g}), and  for head-on collisions where the initial wave functions are uniformly distributed in the transverse direction, the scattering probability following from Eq. (\ref{eq:sptot}) is
\bea
{\prob}_{\text{tot}} = \frac{\sigma_{\text{tot}}}{L^2} + \mathcal{O} (\delta_\tv^2\dtp/|\vec{k}|^3)\ .\label{eq:prcsr}
\eea
This result can be understood intuitively from the definition of scattering cross section \cite{Peskin:1995ev}
\begin{equation}
  \sigma=  \frac{N}{\mathbf{d}_\A \mathbf{l}_\A \mathbf{d}_\B \mathbf{l}_\B {\cal A}} \ ,
\end{equation}
where  $N$ is the expectation value for the number of scattering events, $\mathbf{d}_{\A/\B}$ are the number density of two \textit{uniform} beams,  $\mathbf{l}_{\A/\B}$ are the lengths of the two beams, and ${\cal A}$ is the cross-sectional area of the beams. In our case, we are scattering two  particles so each ``beam'' only contains one particle, with the transverse area ${\cal A}$ of beam given by that of the wave packet: $L^2$. That is $\mathbf{d}_\A \mathbf{l}_\A {\cal A} = \mathbf{d}_\B \mathbf{l}_\B {\cal A} = 1$ and $N = {\prob}$, thereby confirming Eq. (\ref{eq:prcsr}). The same reasoning applies to the elastic/inelastic probabilities in Eq. (\ref{eq:eliegp}) as well. 

To conclude this section, we highlight the fact that taking the plane wave limit leads to a simple interpretation of scattering probabilities as the cross section in unit of the transverse size of the wave packet. We performed a perturbative expansion in the plane wave limit. However, the result holds to all orders in the coupling constants of the QFT.

\section{Bipartite entanglement entropy for ${\rm AB}\to {\rm AB}$}

\label{sec:eetp}

In Section \ref{sec:iswp} we categorized the quantum numbers of each particle into the kinematic data ${\cal H}_{\rm kin}$, labelled by the momenta, and the non-kinematic data ${\cal H}_{\rm f}$ which we call flavors and may include internal quantum numbers or spins. The total Hilbert space ${\cal H}_{\rm AB}$ of the elastic ${\rm AB} \to {\rm AB}$ scattering then consists of
\bea
{\cal H}_{\rm AB} = {\cal H}_{{\rm kin, A}}\otimes {\cal H}_{{\rm f,A}} \otimes {\cal H}_{{\rm kin,B}}\otimes {\cal H}_{{\rm f,B}}\ ,
\eea
where ${\cal H}_{\rm kin,A/B}$ and ${\cal H}_{\rm f,A/B}$ are the  kinematic data and flavor quantum numbers of particle-A/B, respectively. In this section we will compute the subsystem entanglement entropy for different ways of dividing the total Hilbert space ${\cal H}_{\rm AB}$ into bipartite systems. They are

\begin{itemize}
    \item ${\cal H}_{\rm AB} ={\cal H}_{\rm A} \otimes {\cal H}_{\rm B}$, where ${\cal H}_{\rm A/B} = {\cal H}_{{\rm kin, A/B}}\otimes {\cal H}_{{\rm f,A/B}}$. In this case the bipartite system is separated along the particle labels.
     \item ${\cal H}_{\rm AB} ={\cal H}_{\rm kin} \otimes {\cal H}_{\rm f}$, where ${\cal H}_{\rm kin}= {\cal H}_{\rm kin,A}\otimes {\cal H}_{\rm kin,B}$ and ${\cal H}_{\rm f}={\cal H}_{\rm f,A}\otimes {\cal H}_{\rm f,B}$. In this case the bipartite system is divided along kinematic data versus the flavor quantum numbers.
     \item ${\cal H}_{\rm AB} ={\cal H}_{\rm f,A} \otimes \overline{{\cal H}_{\rm f,A}}$, where $\overline{{\cal H}_{\rm f,A}}={\cal H}_{{\rm kin, A}}\otimes {\cal H}_{{\rm kin,B}}\otimes {\cal H}_{{\rm f,B}}$ is the complementarity of ${\cal H}_{\rm f,A}$. One could  consider other possibilities such as  ${\cal H}_{\rm AB} ={\cal H}_{\rm kin,B} \otimes \overline{{\cal H}_{\rm kin,A}}$, which will be commented on later.
\end{itemize}

\subsection{${\cal H}_{\rm AB} ={\cal H}_{\rm A} \otimes {\cal H}_{\rm B}$: Between particle-A and particle-B}

\label{sec:eea}

In this case we consider the quantum  entanglement between the two particles. For the initial state, we trace over the subspace of particle B in Eq. (\ref{eq:dmi}) to get  the following reduced density matrix:
\bea
\rho^{\text{i}}_\A = \sum_{i,j,\bar{i},\bar{j}} \Omega_{i\bar{i}} (\Omega^\dagger)_{\bar{j}j}  |\psi_\A, i \>\<\psi_\A, j| \< \psi_\B, \bar{j}| \psi_\B, \bar{i} \> = \sum_{i,j} (\Omega  \Omega^\dagger)_{ij}  |\psi_\A, i \>\<\psi_\A, j| \ ,
\eea
from which we see
\bea
\tr (\rho^{\text{i}}_\A)^n = \tr (\Omega^\dagger \Omega )^n\ .
\eea
The $n$th order Tsallis  and R\'enyi entropies for the incoming state are
\bea
\mathcal{E}^{\text{i}}_{n,\text{T},\A} = \frac{1- \tr (\Omega^\dagger \Omega )^n}{n-1}\ ,\qquad \mathcal{E}^{\text{i}}_{n,\text{R},\A} = \frac{1}{1-n}\ln \tr   (\Omega^\dagger \Omega )^n\ .\label{eq:eair} \label{eq:eait}
\eea
One special case is when the initial flavor states are not  entangled in the flavor space at all,  $\Omega_{i\bar{i}} = \omega_i \omega'_{\bar{i}}$ where $\omega$ and $\omega'$ specify the flavor of particle A and B, respectively, and $|\omega|^2 = |\omega'|^2 = 1$. Then
\bea
\Omega \Omega^\dagger \Omega  = \Omega, \quad \tr   (\Omega^\dagger \Omega )^n = \tr   (\Omega^\dagger \Omega ) = 1\label{eq:ueicon}
\eea
and the entropy vanishes: $\mathcal{E}^{\text{i}}_{n,\text{T}/\text{R},\A} = 0$.

\begin{figure}[tbp]%
    \centering
    \subfloat[\centering \label{fig:rhoi}]{{\includegraphics[height=0.07\textheight]{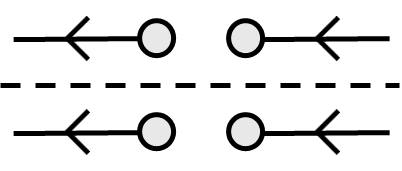} }}%
    \qquad
    \subfloat[\centering \label{fig:rhoia}]{{\includegraphics[height=0.07\textheight]{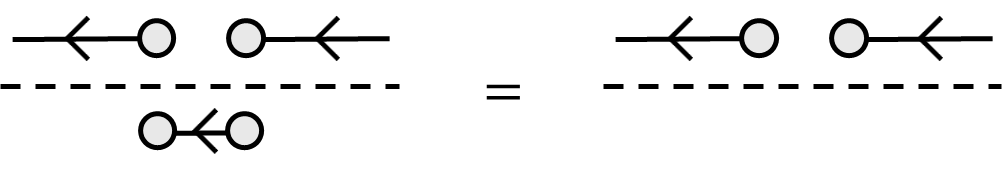} }}%
    \caption{(a) The kinematic data of the initial density matrix $\rho^\text{i}$. The dashed line separate particle-A from particle-B. (b) The kinematic data of the reduced density matrix $\rho^\text{i}_\A$, where $\vec{p}_{\rm B}$ is traced over.}
\end{figure}

For the final state entanglement entropies we need to compute $\tr ( \rho^{\text{f}}_\A )^n$, where $\rho^{\text{f}}_\A = \tr_\B ( \rho^{\text{f}})$ is the final state reduced density matrix, with $\rho^{\text{f}}$ given by Eq. (\ref{eq:fdmpn}). At this stage, it will be convenient to introduce a diagrammatic representation of the kinematic (not flavor) flow of the traces involved, similar to that in Ref. \cite{Cheung:2023hkq}, with the following rules:
\begin{itemize}
\item In each diagram, the upper/lower half separated by the dashed lines represents the kinematic data of particle A/B.

\item There are multiple wave packets involved in evaluating $\tr ( \rho^{\text{f}}_\A )^n$. Each wave packet is represented by a small circle. The total number of wave packets is represented by $n_\psi$.

\item Similarly there are multiple transition matrices $T$ or $T^\dagger$ in the computation, each of which is represented by a grey block and the number of blocks is denoted by $n_T$. Each block comes with a 4-momentum $\delta$-function as in Eq. (\ref{eq:rtm}).

\item  An arrowed line signifies taking the trace over momentum, which we call the momentum flow. An open end represents a momentum that is not traced/contracted and a  connected arrowed line represents a 3-momentum integration. There can be $n_{\int}$ arrowed lines, each end contracted with either a circle (wave packet) or a block (transition matrix). Moreover, an arrowed line connecting  two circles is simply the normalization of the wave packet in Eq. (\ref{eq:wfnrq}) and can be omitted. The arrow keeps track of whether the momentum is in-going or out-going with respect to the transition matrices.
\end{itemize}
For example, the kinematic data in the initial density matrix $\rho^\text{i}$, $|\psi_\A ;\psi_\B\>\<\psi_\A;\psi_\B|$,   is shown in Fig. \ref{fig:rhoi}, and the action of  tracing over particle $B$ to compute $\rho^\text{i}_\A$ is represented in Fig. \ref{fig:rhoia}.

\begin{figure}[tbp]
\centering
\subfloat[\centering \label{fig:rhofa1}]{{\includegraphics[height=0.04\textheight]{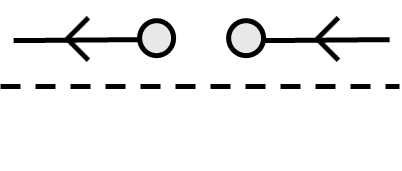} }}%
    \quad
    \subfloat[\centering \label{fig:rhofa2}]{{\includegraphics[height=0.04\textheight]{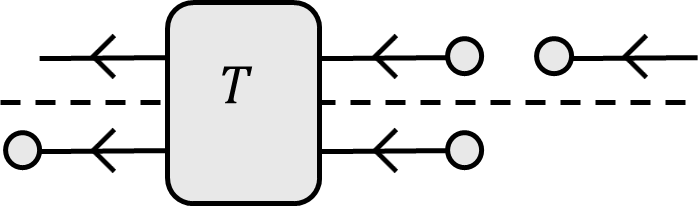} }}
    \quad
    \subfloat[\centering \label{fig:rhofa3}]{{\includegraphics[height=0.04\textheight]{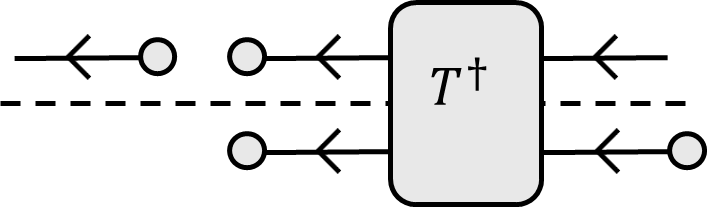} }}
    \quad
    \subfloat[\centering \label{fig:rhofa4}]{{\includegraphics[width=0.26\textwidth]{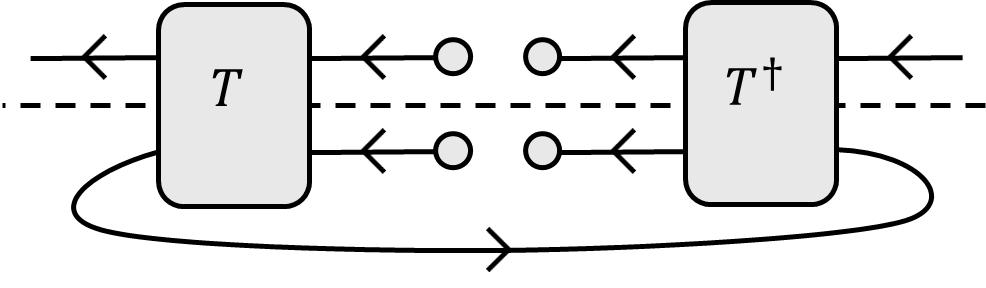} }}
\caption{Contributions to the reduced density matrix $\rho^\text{f}_\A$: (a), (b), (c), and (d) correspond to the first, second, third and fourth term in Eq. (\ref{eq:fourcont}).\label{fig:rhofa}}
\end{figure}

\begin{figure}[tbp]%
    \centering
    \subfloat[\centering \label{fig:ont1}]{{\includegraphics[width=0.2\textwidth]{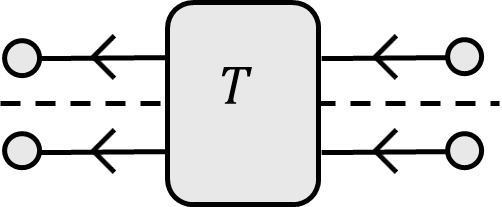} }}%
    \quad
    \subfloat[\centering \label{fig:ont3}]{{\includegraphics[width=0.6\textwidth]{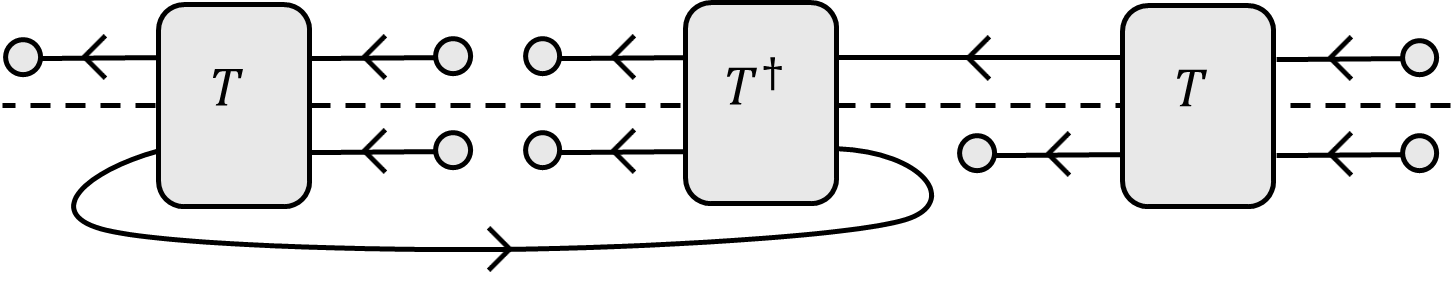} }}%
    \caption{(a) The trace for the product of Figs. \ref{fig:rhofa1} and \ref{fig:rhofa2}. (b) The trace for the product of Figs. \ref{fig:rhofa2} and \ref{fig:rhofa4}.\label{fig:fig5}}%
\end{figure}

For computations involving the final state, the kinematic data of $\rho^{\text{f}}$ contains  4 terms from expanding Eq. (\ref{eq:fdmpn}), shown schematically as
\bea
\label{eq:fourcont}
\rho^{\text{f}} \sim S|\text{in}\> \< \text{in}| S^\dagger = |\text{in}\> \< \text{in}| + iT|\text{in}\> \< \text{in}| - i|\text{in}\> \< \text{in}|T^\dagger +T|\text{in}\> \< \text{in}|T^\dagger\ .
\eea
Tracing over $\{\vec{p}_\B, f_\B \}$ we obtain
the reduced density matrix $\rho^{\text{f}}_\A $, which is shown diagrammatically in Fig. \ref{fig:rhofa}. When computing  $\tr ( \rho^{\text{f}}_\A )^2$, for instance, there are sixteen contributions and the diagrams are   a useful bookkeeping device to keep track of the proliferating terms. For instance, Fig. \ref{fig:ont1} denotes the product of the first two terms in Eq. (\ref{eq:fourcont}), which are obtained by ``contracting''  Figs. \ref{fig:rhofa1} and \ref{fig:rhofa2}, while Fig. \ref{fig:ont3} is the product of Figs. \ref{fig:rhofa2} and \ref{fig:rhofa4}. See Appendix \ref{app:dkd} for a complete list of terms in $\tr ( \rho^{\text{f}}_\A )^2$. Furthermore, from the unitarity of the $S$-matrix,   we have $2\, \text{Im}\, T = T^\dagger T$, thus Fig. \ref{fig:ont1} can also describe $\prob_{\text{tot}} = \< \text{in}| T^\dagger   T |\text{in} \> =2 \< \text{in}| \text{Im}\, T  |\text{in} \>$.  In Appendix \ref{app:dkd} we show a diagrammatic representation of the optical theorem. 


Now, we would like to compute leading contributions in the plane wave limit for quantities appearing in the entropies, i.e. $\tr ( \rho^{\text{f}}_\A )^n$, for which all momenta are contracted. In other words, in diagrams representing $\tr ( \rho^{\text{f}}_\A )^n$ there is no open-ended momentum line -- all momentum lines are contracted either with a wave packet (a small circle) or a transition matrix, $T$ or $T^\dagger$, which is a grey block. This is exemplified in Fig. \ref{fig:fig5}. Next we explain the power counting rules  to identify the leading diagrams  in the plane wave limit.

\begin{figure}[tbp]
\centering
 \subfloat[\centering \label{fig:sot}]{{\includegraphics[width=0.47\textwidth]{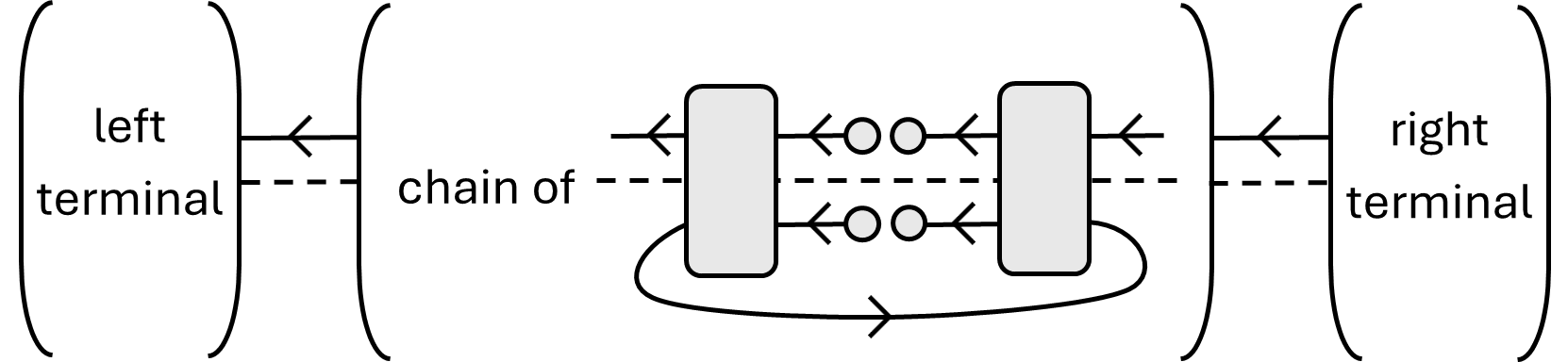} }}%
    \quad
    \subfloat[\centering \label{fig:sotto}]{{\includegraphics[width=0.47\textwidth]{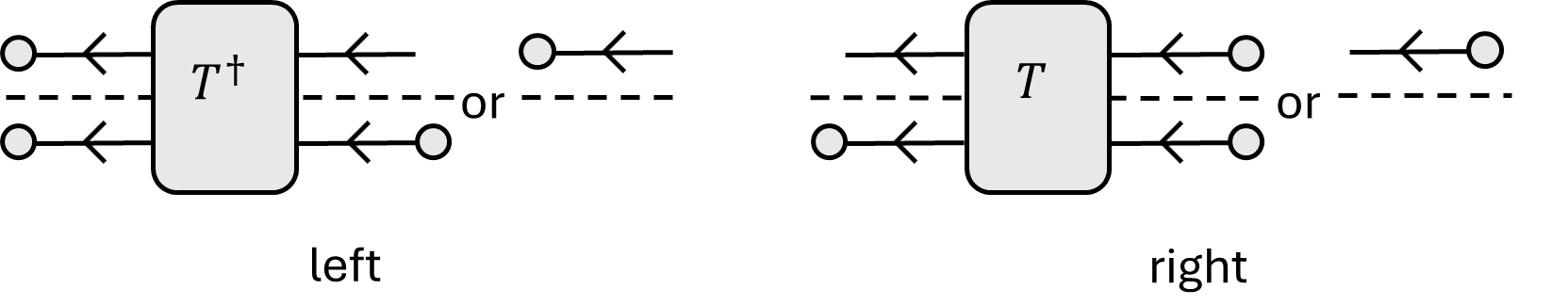} }}%
\caption{(a) The tree diagram, which is a chain of transition matrices that terminates on two ends. (b) The options for the left/right terminals in the tree diagrams.}
\end{figure}

There are two types of diagrams encountered in the computation of $\tr ( \rho^{\text{f}}_\A )^n$. The first type is called the tree diagram, where every momentum flowing through an arrowed line   can be determined entirely from the external momenta in the plane wave limit.\footnote{Recall that in the plane wave limit each wave packet approaches a 3-momentum $\delta$-function. Moreover, each grey box carries a 4-momentum $\delta$-function.} The second type is called the loop diagram where at least one momentum cannot be expressed in terms of external momentum and, therefore, need to be integrated over.

\begin{figure}[tbp]
\centering
    \subfloat[\centering \label{fig:sct}]{{\includegraphics[width=0.45\textwidth]{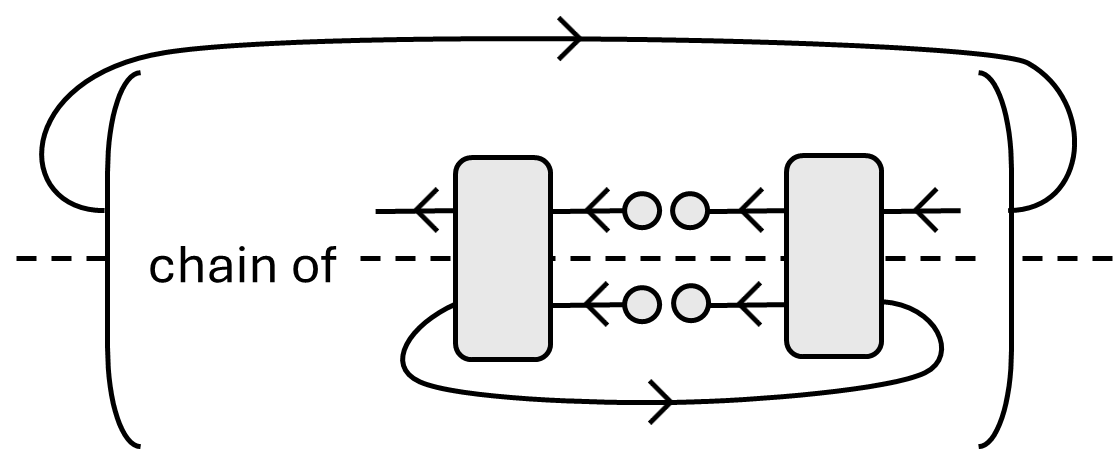} }}%
    \qquad
    \subfloat[\centering \label{fig:cnt4}]{{\includegraphics[width=0.45\textwidth]{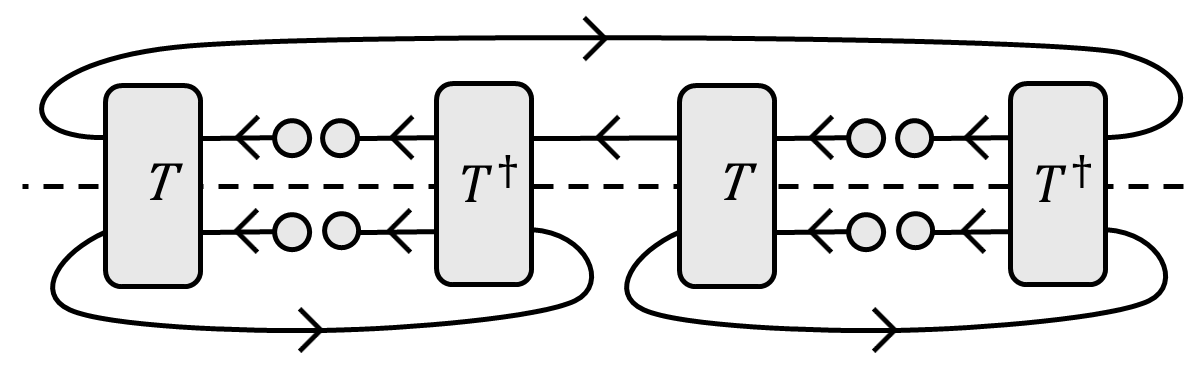} }}%
\caption{(a) The loop diagram, where the transition matrices are connected to form a closed ``loop.'' (b) A loop diagram with $n_T = 4$.}
\end{figure}

The tree diagram is represented in Fig. \ref{fig:sot}, which is a chain of the ``blocks''--the transition matrices--that terminates on two ends, with the possible terminals shown in Fig. \ref{fig:sotto}. Examples include Figs. \ref{fig:ont1} and \ref{fig:ont3} which  appear in $\tr ( \rho^{\text{f}}_\A )^2$.  By inspection one observes that for a tree diagram with  $n_T$ transition matrices,  there are $n_\psi = 2n_T + 2$ wave packets and $n_{\int} = 3n_T+1$ integrations. On the other hand, the loop diagram is shown in Fig. \ref{fig:sct}, which is a chain of blocks that form a closed ``loop.'' In $\tr ( \rho^{\text{f}}_\A )^2$ there is such a term in Fig. \ref{fig:cnt4}, which comes from squaring Fig. \ref{fig:rhofa4}. In these diagrams  $n_T$  is even, and when we compute $\tr ( \rho^{\text{f}}_\A )^n$, they  appear with  $n_T\ge 4$. In addition, there are $n_\psi = 2n_T$ wave packets and $n_{\int} = 3n_T$ 3-momentum integrations.

The notion of tree vs. loop is similar to that in the Feynman diagram: Each grey block, representing a transition matrix $T$ or $T^\dagger$, is a ``quartic vertex.'' Each arrowed line, if attached to the grey blocks on both ends, is an ``internal line.''  If the line is attached to a grey block on one side and to a circle, i.e. wave packet, on the other side, it is an ``external leg.'' Therefore, $n_T$ is the number of quartic vertices, $n_\psi$ is the number of external legs, and $n_{\int} - n_\psi$ is the number of internal lines. The number of loops is then given by
\bea
n_L =(n_{\int} - n_\psi) - (n_T-1) =n_{\int} - n_\psi- n_T + 1 .\label{eq:loopc}
\eea
For tree diagrams,  $n_\psi = 2n_T + 2$ and $n_{\int} = 3n_T+1$, thus we have $n_L = 0$. For loop diagrams, $n_L=1$ since $n_\psi = 2n_T $ and $n_{\int} = 3n_T$. We emphasize that the ``loop integration'' in these diagrams are simply momentum integrations that are not completely fixed by 3-momentum delta functions. Such momentum integrations also appear in the derivation of optical theorem which  allow us to go from the probability (i.e., amplitude-squared) to the cross-section. In this sense they are really the phase space integrations.

In the plane wave limit, the momentum in any ``external leg'' will collapse to $\vec{k}_{\A/\B}$, as the wave packet becomes infinitely peaked. Given that each grey block--transition matrices $T$ or $T^\dagger$--will contribute a 4-momentum $\delta$-function,  in a tree diagram the momentum flowing through the ``internal line'' is completely fixed in the plane wave limit and expressed in terms of $\vec{k}_{\A/\B}$. After the 4-momentum $\delta$-functions are integrated over, the residual momentum integrations are those which have support only in the small region inside the wave packet and admit the scaling behavior given by Eq. (\ref{eq:sb1di}). Thus the scaling behavior of the tree diagrams can be directly generalized from the discussion of $I_0 (|\vec{k}|)$ in Sec. \ref{sec:fses}:
\bea
 && \Big( \int dp_\tv \Big)^{2n_{\int}} \Big( \int dp_z \Big)^{n_{\int}} \psi^{n_\psi} \, \left[\delta^{4} \Big(\sum p \Big) \right]^{n_T} 
 \sim \Big( \int dp_\tv \Big)^{4n_T+2} \Big( \int dp_z \Big)^{n_T+1} \psi^{2n_T+2}\non\\
 & \sim & \delta_{\tv}^{4n_T+2} \dtp^{n_T + 1} \frac{1}{\delta_\tv^{2n_T + 2} \dtp^{n_T + 1 }} \sim \delta_{\tv}^{2n_T}.
\eea
On the other hand, in the loop diagram there are not enough wave packets and transition matrices to fix the momentum in the internal lines. In this case we encounter the following integral, 
\bea
\int_{q_1} \int_{q_2} \delta^4 (k_{\A} + k_{\B} - q_1 - q_2) \sim \left.\int_{q_1} \delta (E_{k_\A} + E_{k_\B} - E_{q_1} - E_{q_2})\right|_{\vec{q}_2 = \vec{k}_\A + \vec{k}_\B - \vec{q}_1}.\label{eq:gpsi}
\eea
So while $|\vec{q}_{1,2}|$ are fixed to be $|\vec{k}|$ by the 4-momentum $\delta$-functions, there is still the angular integration to be performed over the entire phase space, which is  not suppressed in the plane wave limit. So each $(\int_p)^2 \delta^4(\sum p)$ of the above form will be enhanced over the tree diagrams by $1/\delta_\tv^2$. Thus the scaling behavior of the loop diagram is given by
\bea
&& \Big( \int dp_\tv \Big)^{2(n_{\int}-2)} \Big( \int dp_z \Big)^{n_{\int}-2} \psi^{n_\psi} \, \left[\delta^{4} \Big(\sum p \Big) \right]^{n_T-1}  \sim \Big( \int dp_\tv \Big)^{4n_T-2} \Big( \int dp_z \Big)^{n_T} \psi^{2n_T}\non\\
 & \sim & \delta_{\tv}^{4n_T-2} \dtp^{n_T } \frac{1}{\delta_\tv^{2n_T } \dtp^{n_T  }} \sim \delta_{\tv}^{2n_T-2}.\label{eq:sbcd}
\eea
This analysis shows  the leading scaling behavior in the plane wave limit for both the tree and loop diagrams only depend on  $\delta_\tv$; the tree diagrams start at $\ordr (\delta_\tv^2)$ with $n_T=1$, while the loop diagrams only start contributing at $\ordr (\delta_\tv^6)$ with $n_T = 4$. We conclude that the leading contribution in the plane wave limit is given by the  $n_T$ = 1 diagram in Fig. \ref{fig:ont1} and its conjugate. It is worth stressing that this result is obtained by merely taking the plane wave limit, which does not involve any perturbative expansion in the coupling constants. As such  it is  valid to all orders in the coupling strength.

From Eq. (\ref{eq:fdmpn}) and taking the plane wave limit, we thus arrive at
\bea
\tr\ ( \rho^{\text{f}}_\A )^n
 &=&  \tr \left(\Omega^\dagger \Omega \right)^n \left(1 + n\prob_{\inel}\right) - \frac{n\,I_0 (|\vec{k}|) }{2|\vec{k}|\sqrt{s}} \left\{ \text{Im} \ \tr \left[ (\Omega^\dagger \Omega)^{n-1} \Omega^\dagger M^{\text{F}} (\Omega)   \right]  + \ordr (\dtp/|\vec{k}|) \right\} \non\\
&=&  \tr \left(\Omega^\dagger \Omega \right)^n - \frac{n\,I_0 (|\vec{k}|) }{2  |\vec{k}|\sqrt{s} } \left\{\text{Im} \ \tr \left[( \Omega^\dagger \Omega )^{n-1}\Omega^\dagger M^{\text{F}} (\Omega)   \right] \right.   \non\\
&& \qquad \qquad \left.- 2 |\vec{k}|\sqrt{s} \ \sigma_{\inel}\ \tr (\Omega^\dagger \Omega )^n \right\}  + \ordr (\delta^2_\tv\dtp/|\vec{k}|^3)\ ,\label{eq:rhonc}
\eea
where $\left[ M^{\text{F}} (\Omega) \right]_{i\bar{i}} \equiv  \sum_{j,\bar{j}} \Omega_{j\bar{j}}   M_{j\bar{j},i\bar{i}}^\text{F}$ and $M_{i\bar{i},j\bar{j}}^{\text{F}} = M_{i\bar{i},j\bar{j}} (k_\A, k_\B; k_\A, k_\B)$
is the forward scattering amplitude for the flavor configuration  $i\bar{i} \to j \bar{j}$. Recall in Eqs. (\ref{eq:eliegp}) and (\ref{eq:i0gsc}) we have shown that $P_\inel  \approx I_0 (|\vec{k}|) \, \sigma_\inel$ is $\ordr (\delta_\tv^2/|\vec{k}|^2)$. Thus in the first line of Eq. (\ref{eq:rhonc}) we have expanded the normalization factor $1/(1-\prob_{\inel})$ in Eq. (\ref{eq:fdmpn}) to the leading order. 

In Eq. (\ref{eq:rhonc}) all the information of the initial wave packet  is contained in the factor $I_0 (|\vec{k}|) $, which we have already analyzed thoroughly in Sec. \ref{sec:fses}. The Tsallis and  R\'enyi entropies are then given by 
\bea
\mathcal{E}^{\text{f}}_{n,\text{T},\A} &=& \frac{1}{n-1}\left\{1- \tr (\Omega^\dagger \Omega )^n + \frac{nI_0 (|\vec{k}|) }{2  |\vec{k}|\sqrt{s} } \bigl[\text{Im} \ \tr \left(( \Omega^\dagger \Omega )^{n-1}\Omega^\dagger M^{\text{F}} (\Omega)   \right) \right.   \non\\
&&\left. \phantom{\frac{nI_0 (|\vec{k}|) }{2  |\vec{k}|\sqrt{s} }}- 2 |\vec{k}|\sqrt{s}\  \sigma_{\inel}\ \tr (\Omega^\dagger \Omega )^n  \right\}+ \ordr (\delta^2_\tv\dtp/|\vec{k}|^3),\label{eq:fseta}\\
\mathcal{E}^{\text{f}}_{n,\text{R},\A} &=& \frac{1}{1-n}\left\{\log \tr (\Omega^\dagger \Omega )^n - \frac{nI_0 (|\vec{k}|) }{2  |\vec{k}|\sqrt{s} \ \tr (\Omega^\dagger \Omega )^n} \bigl[\text{Im} \ \tr \left(( \Omega^\dagger \Omega )^{n-1}\Omega^\dagger M^{\text{F}} (\Omega)   \right) \right.   \non\\
&&\left. \phantom{\frac{nI_0 (|\vec{k}|) }{2  |\vec{k}|\sqrt{s} }}- 2 |\vec{k}|\sqrt{s}\  \sigma_{\inel}\ \tr (\Omega^\dagger \Omega )^n \bigr] \right\}  + \ordr (\delta^2_\tv\dtp/|\vec{k}|^3)\label{eq:fsetr}.
\eea
For the R\'enyi entropy in the above, we have expanded the argument of the logarithm in Eq. (\ref{eq:etrdef}) using the plane wave limit as well. The following term in the entropies,
\bea
 \tr [ ( \Omega^\dagger \Omega )^{n-1}\Omega^\dagger M^{\text{F}} (\Omega)  ] = \sqrt{\tr (\Omega^\dagger \Omega)^{2n-1}} \sum_{i,\bar{i},j, \bar{j}} \Omega_{i\bar{i}} M_{i\bar{i},j\bar{j}}^{\text{F}} \left(\Omega_{(n-1)}^\dagger\right)_{\bar{j} j},\label{eq:gott}
 \eea
 is the amplitude with forward kinematics where the initial flavor configuration is $\Omega$ and the properly normalized final flavor configuration is $\Omega_{(n-1)}$, with
 \bea
 \Omega_{(n-1)} = \frac{\Omega (\Omega^\dagger \Omega)^{n-1}}{\sqrt{\tr (\Omega^\dagger \Omega)^{2n-1}}}.
 \eea
The change in the Tsallis entropy between the initial and final states is given by
\bea
\Delta \mathcal{E}_{n,\text{T},\A}  \equiv  \mathcal{E}^{\text{f}}_{n,\text{T},\A} - \mathcal{E}^{\text{i}}_{n,\text{T},\A} &=&  \frac{nI_0 (|\vec{k}|) }{2(n-1)  |\vec{k}|\sqrt{s} } \bigl[\text{Im} \ \tr \left(( \Omega^\dagger \Omega )^{n-1}\Omega^\dagger M^{\text{F}} (\Omega)   \right)   \non\\
&&- 2 |\vec{k}|\sqrt{s} \ \sigma_{\inel}\ \tr (\Omega^\dagger \Omega )^n \bigr]  + \ordr (\delta^2_\tv\dtp/|\vec{k}|^3),
\eea
which is proportional to $I_0 (|\vec{k}|)$. For the change in the R\'enyi entropy we have
\bea
\Delta \mathcal{E}_{n,\text{R},\A}  \equiv  \mathcal{E}^{\text{f}}_{n,\text{R},\A} - \mathcal{E}^{\text{i}}_{n,\text{R},\A} =\frac{\Delta \mathcal{E}_{n,\text{T},\A}}{\tr (\Omega^\dagger \Omega )^n} + \ordr (\delta^2_\tv\dtp/|\vec{k}|^3).
\eea

Now consider the case of unentangled initial states, where Eq. (\ref{eq:ueicon}) converts Eq. (\ref{eq:gott}) into
\bea
 \tr [ ( \Omega^\dagger \Omega )^{n-1}\Omega^\dagger M^{\text{F}} (\Omega)  ]= \tr [\Omega^\dagger M^{\text{F}} (\Omega)] 
= \sum_{i,\bar{i},j, \bar{j}} \Omega_{i\bar{i}} M_{i\bar{i},j\bar{j}}^{\text{F}} \left(\Omega^\dagger\right)_{\bar{j} j},\label{eq:gotues}
\eea
which is actually the forward amplitude when both the initial and final state flavor configurations are given by $\Omega$, i.e. the final state flavor does not change. Then the optical theorem gives
\bea
\text{Im} \  \tr [\Omega^\dagger M^{\text{F}} (\Omega)] = 2|\vec{k}|\sqrt{s} \ \sigma_{\text{tot}} \label{eq:ot}
\eea
and
\bea
\mathcal{E}^{\text{f}}_{n,\text{T}/\text{R},\A}   &=& \frac{n}{n-1} I_0 (|\vec{k}|) (\sigma_{\text{tot}} -  \sigma_{\inel}   ) + \ordr (\delta^2_\tv\dtp/|\vec{k}|^3)  \non\\
&=& \frac{n}{n-1} I_0 (|\vec{k}|) \, \sigma_{\el} + \ordr (\delta^2_\tv\dtp/|\vec{k}|^3) .\label{eq:eadsw} 
\eea
In other words, the entanglement entropies are proportional to the elastic cross sections in the plane wave limit. Notice that  $\sigma_{\text{tot}}$ in the above comes from the numerator of Eq. (\ref{eq:fdmpn}), i.e. Fig. \ref{fig:ont1} and its conjugate, while $\sigma_{\inel}$ comes from expanding the overall normalization factor $1/(1-\prob_\inel)$ in Eq. (\ref{eq:fdmpn}): They nontrivially conspire to give $\sigma_\el$. We do not have such a conspiracy   for general, entangled initial states, as shown in Eqs. (\ref{eq:fseta}) and (\ref{eq:fsetr}). It is also important to identify from Eq. (\ref{eq:eliegp}) that
\bea
\mathcal{E}^{\text{f}}_{n,\text{T}/\text{R},\A}  = \frac{n}{n-1} \prob_\el  + \ordr (\delta^2_\tv\dtp/|\vec{k}|^3),
\eea
i.e. the final state entanglement entropy is exactly the elastic scattering probability at the leading order in the plane wave limit when the initial states are not entangled, up to a simple numerical factor. This is a conclusion that is independent of the details of the wave packets, and the discussion on examples of wave packet configurations in Sec. \ref{sec:fses} directly carries over to the entanglement entropies presented here.

\subsection{${\cal H}_{\rm AB} ={\cal H}_{\rm kin} \otimes {\cal H}_{\rm f}$: Between flavor and momentum}
\label{sec:efm}

\begin{figure}[tbp]
\centering
\includegraphics[width=0.55\textwidth]{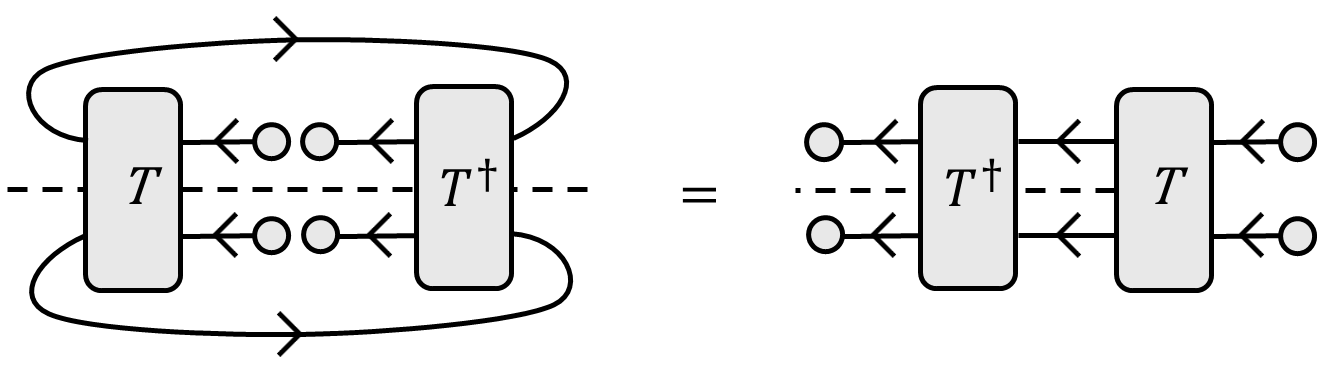}
\caption{The term that generates $M^{(2)}$, which is a loop diagram with $n_T = 2$.\label{fig:cnt2}}
\end{figure}

Now consider the entanglement entropy between the flavor space and the momentum space, ${\cal H}_{\rm AB} ={\cal H}_{\rm kin} \otimes {\cal H}_{\rm f}$.\footnote{When this work was in progress,  Ref. \cite{Kowalska:2024kbs} appeared which also considers various configurations for the entanglement entropy that is not between the two particles.} Tracing over the momentum subspace of Eqs. (\ref{eq:dmi}) and (\ref{eq:fdmpn}), the reduced density matrices for the inital and final states are computed to be
\bea
\left(\rho^{\text{i}}_f\right)_{i\bar{i}, j\bar{j}} &=& \Omega_{i \bar{i}} \Omega^*_{j \bar{j}}, \\
 \left(\rho^{\text{f}}_f\right)_{i\bar{i}, j\bar{j}} &=& \Omega_{i \bar{i}} \Omega^*_{j \bar{j}} ( 1 + \prob_\inel ) + \frac{I_0 (|\vec{k}|)}{4 |\vec{k}| \sqrt{s} }\Big(i \left[M^{\text{F}} (\Omega) \right]_{i \bar{i}}  \Omega^*_{j \bar{j}} - i\Omega_{i \bar{i}} \left[M^{\text{F}} (\Omega) \right]_{ \bar{j}j} ^*    + \left[M^{(2)} (\Omega) \right]_{i \bar{i}, j \bar{j}} \Big) \non\\
&&+ \ordr (\delta^2_\tv\dtp/|\vec{k}|^3),\label{eq:rdmff}
\eea
where
\bea
\left[M^{(2)} (\Omega) \right]_{i \bar{i}, j \bar{j}}&\equiv& \sum_{a, \bar{a},b,\bar{b}} \int_{ p,q}  \frac{1}{\sqrt{4E_{p} E_{q}}}\Omega_{a \bar{a}} \Omega^*_{b \bar{b}}    M_{a \bar{a}, i \bar{i}} (k_\A, k_\B; p,q) M_{b \bar{b}, j \bar{j}}^* (k_\A, k_\B; p,q) \non\\
&&\times (2 \pi)^4 \delta^{4} (p+q - k_\A - k_\B)  .\label{eq:m2def}
\eea
Here we compute the reduced density matrix by tracing over the momentum, and this is the step where we take the plane wave limit. The resulting reduced density matrices are discrete and independent of the momentum. The terms in Eq. (\ref{eq:rdmff}) involving $M^\text{F}$ comes from Fig. \ref{fig:ont1} and its conjugate, while the $M^{(2)}$ term corresponds to Fig. \ref{fig:cnt2}, which is a loop  diagram with $n_T = 2$ transition matrices. From Eq. (\ref{eq:sbcd}) we see that its scaling behavior is $\delta_\tv^2$, which is the same as Fig. \ref{fig:ont1}.  Actually, Eq. (\ref{eq:m2def}) offers a simple example of the phase space integral described in Eq. (\ref{eq:gpsi}). Notice that in Sec. \ref{sec:eea}, when the entanglement entropy between the two particles are computed, all loop diagrams have $n_T \ge 4$, while here an $n_T = 2$ loop diagram appears. Also notice that the diagram in Fig. \ref{fig:cnt2} can describe $\prob_\el = \<\text{in} | T^\dagger P_{\A\B} T | \text{in} \>$ computed in Sec. \ref{sec:fses}, confirming that it should scale as $\delta_\tv^2$.

For the initial states, $\tr \left(\rho^{\text{i}}_f\right)^n = 1$ and  the entanglement entropy computed from Eq. (\ref{eq:etrdef}) is $\mathcal{E}^{\text{i}}_{n,\text{T},f} = \mathcal{E}^{\text{i}}_{n,\text{R},f} =0$, as the flavor state  is  not entangled with the momentum state. For the final state in the plane wave limit, we have
\bea
\tr \left(\rho^{\text{f}}_f\right)^n &=& 1- \frac{nI_0 (|\vec{k}|)}{4 |\vec{k}|\sqrt{s} } \Big[2\ \text{Im} \ \tr \left(  \Omega^\dagger M^{\text{F}} (\Omega)   \right)  -4|\vec{k}| \sqrt{s}\  \sigma_{\inel}  - \sum_{i ,\bar{i}, j, \bar{j} } \Omega^*_{i \bar{i}} \Omega_{j \bar{j}} \left[M^{(2)} (\Omega) \right]_{i \bar{i}, j \bar{j}}   \Big] \non\\&&+\ordr (\delta^2_\tv\dtp/|\vec{k}|^3)\non\\
&=& 1 - nI_0 (|\vec{k}|) \left( \sigma_{\text{tot}}- \sigma_\inel - \sigma_{\el,\text{fp}} \right)+ \ordr (\delta^2_\tv\dtp/|\vec{k}|^3)\non\\
&=& 1-nI_0 (|\vec{k}|) \, \sigma_{\el,\text{fc}} + \ordr (\delta^2_\tv\dtp/|\vec{k}|^3)
\eea
where we have used the optical theorem given by Eq. (\ref{eq:ot}),\footnote{Notice that the optical theorem, Eq. (\ref{eq:ot}), holds whether or not the initial flavor states are entangled.} and
\bea
\sigma_{\el,\text{fc}} = \sigma_{\el} - \sigma_{\el,\text{fp}}
\eea
is the semi-inclusive cross section for elastic scattering with at least one of the particles changing flavor. The flavor-preserving cross section is given by
\bea
\sigma_{\el,\text{fp}} &=&\frac{1}{4 |\vec{k}| \sqrt{s}}  \sum_{i ,\bar{i}, j, \bar{j} } \Omega^*_{i \bar{i}} \Omega_{j \bar{j}} \left[M^{(2)} (\Omega) \right]_{i \bar{i}, j \bar{j}}\non\\
&=&\frac{1}{4 |\vec{k} | \sqrt{s}}   \int_{p,q} \frac{(2\pi)^4}{\sqrt{4E_{p} E_{q}}} \delta^4 \left(k_\A + k_\B - p-q\right) \Big| \sum_{i,\bar{i},j,\bar{j}}  \Omega_{i \bar{i}} (\Omega^\dagger)_{ \bar{j}j} M_{i\bar{i}, j\bar{j}} (k_\A,k_\B;p,q) \Big|^2 .
\eea
Therefore, the entanglement entropy is
\bea
\mathcal{E}^{\text{f}}_{n,\text{T}/\text{R},f}= \frac{n}{n-1}I_0 (|\vec{k}|)\, \sigma_{\el,\text{fc}} + \ordr (\delta^2_\tv\dtp/|\vec{k}|^3).
\eea
Notice that as the flavor subspace and momentum subspace are not entangled in our initial state, we can already express the leading order, final state entanglement entropies in terms of a cross section, regardless of whether the initial flavors of the two particles are entangled or not. In this case the entropy selects not only  the elastic cross section, but also certain flavor configurations in the final state. Furthermore, similar to what we have computed in Eq. (\ref{eq:sptot}), it is straightforward to derive the probability for elastic scattering with at least one of the particle changing flavor to be
\bea
\prob_{\el,\text{fc}} =  I_0 (|\vec{k}|)\, \sigma_{\el,\text{fc}} + \ordr (\delta^2_\tv\dtp/|\vec{k}|^3),
\eea
thus we have
\bea
\mathcal{E}^{\text{f}}_{n,\text{T}/\text{R},f}= \frac{n}{n-1}\prob_{\el,\text{fc}}  + \ordr (\delta^2_\tv\dtp/|\vec{k}|^3).
\eea
Once again, we identify the entanglement entropy with a probability at the leading order  in the plane wave limit.

\subsection{${\cal H}_{\rm AB} ={\cal H}_{\rm f,A} \otimes \overline{{\cal H}_{\rm f,A}}$: For the flavor of a single particle}

Next we consider ${\cal H}_{\rm AB} ={\cal H}_{\rm f,A} \otimes \overline{{\cal H}_{\rm f,A}}$ and tracing out the flavor space of particle B as well as  the entire momentum subspace. The initial reduced density matrix is $\rho^{\text{i}}_{f,\A} = \Omega \Omega^\dagger$ and final reduced density matrix is
\bea
\left[\rho^{\text{f}}_{f,\A}\right]_{ij} &=& \left[\Omega \Omega^\dagger\right]_{ij} ( 1 + \prob_{\inel} )  +  \frac{I_0 (|\vec{k}|)}{4 |\vec{k}| \sqrt{s}} \Big( i \left[\tilde{M} (\Omega) \Omega^\dagger \right]_{ij}   - i \left[\Omega \tilde{M}^\dagger (\Omega) \right]_{ij} \non\\
&&+ \sum_{\bar{i}} \left[M^{(2)}  (\Omega) \right]_{i \bar{i}, j \bar{i}} \Big) + \ordr (\delta^2_\tv\dtp/|\vec{k}|^3).
\eea
Similar to what we have seen in Sec. \ref{sec:efm}, the reduced density matrices are discrete. The initial entanglement entropies are
\bea
\mathcal{E}^{\text{i}}_{n,\text{T},f,\A} = \frac{1- \tr (\Omega^\dagger \Omega )^n}{n-1},\qquad \mathcal{E}^{\text{i}}_{n,\text{R},f,\A} = \frac{1}{1-n}\log \tr   (\Omega^\dagger \Omega )^n,
\eea
which are the same as in Sec. \ref{sec:eea}, and will vanish when the initial flavor states are not entangled.

For the final state entropies, we have
\bea
\tr \left(\rho^{\text{f}}_{f,\A}\right)^n &=& \tr (\Omega^\dagger \Omega )^n  - \frac{nI_0 (|\vec{k}|)}{4  |\vec{k}|\sqrt{s}} \Big[2\ \text{Im} \ \tr \left(  (\Omega^\dagger \Omega)^{n-1} \Omega^\dagger M^{\text{F}} (\Omega)   \right) - 4|\vec{k}|\sqrt{s} \ \sigma_{\inel}\ \tr (\Omega^\dagger \Omega )^n \non\\
&& - \sum_{i,j,\bar{i} } \left[(\Omega \Omega^\dagger)^{n-1} \right]_{ij} \left[M^{(2)} (\Omega) \right]_{j \bar{i},  i \bar{i}} \Big] + \ordr (\delta^2_\tv\dtp/|\vec{k}|^3),
\eea
and thus 
\bea
\mathcal{E}^{\text{f}}_{n,\text{T},f,\A} &=& \frac{1}{n-1}\left(1- \tr (\Omega^\dagger \Omega )^n + \frac{nI_0 (|\vec{k}|) }{2  |\vec{k}|\sqrt{s} } \Big[2\, \text{Im} \ \tr \left(( \Omega^\dagger \Omega )^{n-1}\Omega^\dagger M^{\text{F}} (\Omega)   \right)  \right.   \non\\
&&\left. \phantom{\frac{nI_0 (|\vec{k}|) }{2  |\vec{k}|\sqrt{s} }} - 4|\vec{k}|\sqrt{s} \ \sigma_{\inel}\ \tr (\Omega^\dagger \Omega )^n - \sum_{i,j,\bar{i} } \left[(\Omega \Omega^\dagger)^n \right]_{ij} \left[M^{(2)} (\Omega) \right]_{j \bar{i},  i \bar{i}} \Big] \right)\non\\
&&+ \ordr (\delta^2_\tv\dtp/|\vec{k}|^3),\\
\mathcal{E}^{\text{f}}_{n,\text{R},f,\A} &=& \frac{1}{1-n}\left(\ln \tr (\Omega^\dagger \Omega )^n - \frac{nI_0 (|\vec{k}|) }{2  |\vec{k}|\sqrt{s} \ \tr (\Omega^\dagger \Omega )^n} \Big[2\, \text{Im} \ \tr \left(( \Omega^\dagger \Omega )^{n-1}\Omega^\dagger M^{\text{F}} (\Omega)   \right)  \right.   \non\\
&&\left. \phantom{\frac{nI_0 (|\vec{k}|) }{2  |\vec{k}|\sqrt{s} }} - 4|\vec{k}|\sqrt{s} \ \sigma_{\inel}\ \tr (\Omega^\dagger \Omega )^n - \sum_{i,j,\bar{i} } \left[(\Omega \Omega^\dagger)^n \right]_{ij} \left[M^{(2)} (\Omega) \right]_{j \bar{i},  i \bar{i}} \Big] \right)\non\\
&&+ \ordr (\delta^2_\tv\dtp/|\vec{k}|^3).
\eea
The changes of the entanglement entropies are
\bea
\Delta \mathcal{E}_{n,\text{T},f,\A}  &\equiv&  \mathcal{E}^{\text{f}}_{n,\text{T},f,\A} - \mathcal{E}^{\text{f}}_{n,\text{T},f,\A}\non\\
 &=&  \frac{nI_0 (|\vec{k}|) }{4(n-1)  |\vec{k}|\sqrt{s} } \Big[2\, \text{Im} \ \tr \left(( \Omega^\dagger \Omega )^{n-1}\Omega^\dagger M^{\text{F}} (\Omega)   \right)     \non\\
&& - 4|\vec{k}|\sqrt{s} \ \sigma_{\inel}\ \tr (\Omega^\dagger \Omega )^n - \sum_{i,j,\bar{i} } \left[(\Omega \Omega^\dagger)^n \right]_{ij} \left[M^{(2)} (\Omega) \right]_{j \bar{i},  i \bar{i}} \Big]  + \ordr (\delta^2_\tv\dtp/|\vec{k}|^3),\ \\
\Delta \mathcal{E}_{n,\text{R},f,\A}  &\equiv&  \mathcal{E}^{\text{f}}_{n,\text{R},f,\A} - \mathcal{E}^{\text{f}}_{n,\text{R},f,\A} =\frac{\Delta \mathcal{E}_{n,\text{T},f,\A}}{\tr (\Omega^\dagger \Omega )^n} + \ordr (\delta^2_\tv\dtp/|\vec{k}|^3),
\eea
again $\ordr (\delta_\tv^2/|\vec{k}|^2)$ quantities.

For unentangled initial flavor states, recall Eq. (\ref{eq:ueicon}) as well as $\Omega_{i \bar{i}} = \omega_i \omega'_{\bar{i}}$, and we obtain
\bea
\sum_{i,j,\bar{i} } \left[(\Omega \Omega^\dagger)^n \right]_{ij} \left[M^{(2)} (\Omega) \right]_{j \bar{i},  i \bar{i}} =\sum_{i,j,\bar{i} } \omega_i \omega^*_j \left[M^{(2)} (\Omega) \right]_{j \bar{i},  i \bar{i}} = 4 |\vec{k}| \sqrt{s} \ \sigma_{\el,\text{fp(A)}}
\eea
where
\bea
\sigma_{\el,\text{fp(A)}} = \frac{1}{4 |\vec{k} | \sqrt{s}} \sum_{\bar{j}} \int_{p,q}   \ \frac{(2\pi)^4}{\sqrt{4E_{p} E_{q}}} \delta^4 \left(k_\A + k_\B - p-q\right) \Big| \sum_{i,j,\bar{i}} \Omega_{i \bar{i}} \, \omega_j^* M_{i\bar{i}, j\bar{j}} (k_\A,k_\B;p,q) \Big|^2 \ \ 
\eea
is the elastic cross section that preserves the flavor of particle A.
Combining with Eqs. (\ref{eq:gotues}) and (\ref{eq:ot}), we evaluate the entropy in the plane wave limit to be
\bea
\mathcal{E}^{\text{f}}_{n,\text{T}/\text{R},f,\A} =  \frac{n}{n-1}  I_0 (|\vec{k}|)\, \sigma_{\el,\text{fc(A)}} + \ordr (\delta^2_\tv\dtp/|\vec{k}|^3), 
\eea
where
\bea
\sigma_{\el,\text{fc(A)}} = \sigma_{\el} - \sigma_{\el,\text{fp(A)}}
\eea
is the elastic cross section with the type-A particle changing flavor. One can again compute the elastic scattering probability for particle A changing flavor to be
\bea
\prob_{\el,\text{fc(A)}} =  I_0 (|\vec{k}|)\, \sigma_{\el,\text{fc(A)}} + \ordr (\delta^2_\tv\dtp/|\vec{k}|^3),
\eea
and identify the leading order entanglement entropy with scattering probability:
\bea
\mathcal{E}^{\text{f}}_{n,\text{T},\text{fc(A)}}= \frac{n}{n-1}\prob_{\el,\text{fc(A)}}  + \ordr (\delta^2_\tv\dtp/|\vec{k}|^3).
\eea

\subsection{Other partitions}

Given the total Hilbert space in 2-to-2 scattering: ${\cal H}_{\rm AB} = {\cal H}_{{\rm kin, A}}\otimes {\cal H}_{{\rm f,A}} \otimes {\cal H}_{{\rm kin,B}}\otimes {\cal H}_{{\rm f,B}}$, there are more possibilities of constructing a bipartite system. However, these other possibilities lead to the entropies being expressed as cross sections which have already been discussed.  For completeness, below we briefly list the entanglement entropy for other partitions of the system.

For ${\cal H}_{\rm I}= {\cal H}_{{\rm f,A}} \otimes {\cal H}_{{\rm kin,B}}$, which is the flavor of A-particle and momentum of B-particle, and ${\cal H}_{\rm AB}={\cal H}_{\rm I}\otimes \overline{{\cal H}_{\rm I}}$, the initial entanglement entropies $\mathcal{E}^{\text{i}}_{n,\text{T}/\text{R},\times}$ exactly agree with Eq. (\ref{eq:eait}). For the final states, the entanglement entropy $\mathcal{E}^{\text{f}}_{n,\text{T}/\text{R},\times}$ agrees with $\mathcal{E}^{\text{f}}_{n,\text{T}/\text{R},\A}$ given in Eqs. (\ref{eq:fseta}) and (\ref{eq:fsetr}) at the leading order. Therefore, for unentangled initial flavors,
\bea
\mathcal{E}^{\text{f}}_{n,\text{T}/\text{R},\times} = \frac{n}{n-1} I_0 (|\vec{k}|)\,  \sigma_{\el} + \ordr (\delta^2_\tv\dtp/|\vec{k}|^3) = \frac{n}{n-1} \prob_{\el}   + \ordr (\delta^2_\tv\dtp/|\vec{k}|^3) .
\eea

For ${\cal H}_{\rm I}= {\cal H}_{{\rm kin,A}}$, which is just the momentum of A-particle, and ${\cal H}_{\rm AB}={\cal H}_{\rm I}\otimes \overline{{\cal H}_{\rm I}}$, the initial entanglement entropy vanishes. The final entanglement entropy is
\bea
\mathcal{E}^{\text{f}}_{n,\text{T}/\text{R},p,\A}  &=&  \frac{nI_0 (|\vec{k}|)}{2 |\vec{k}|\sqrt{s} } \Big[ \text{Im} \ \tr \left(  \Omega^\dagger M^{\text{F}} (\Omega)   \right)  -2|\vec{k}| \sqrt{s} \sigma_{\inel} \Big] + \ordr (\delta^2_\tv\dtp/|\vec{k}|^3)\non\\
& =& \frac{n}{n-1} I_0 (|\vec{k}|)\,  \sigma_{\el} + \ordr (\delta^2_\tv\dtp/|\vec{k}|^3) = \frac{n}{n-1} \prob_{\el}   + \ordr (\delta^2_\tv\dtp/|\vec{k}|^3) ,
\eea
for any initial flavor configuration.

In Table \ref{tab:sum} we summarize the correspondence between final entanglement entropies and cross sections/probabilities for different kinds of partitions of the system, as well as whether unentangled initial flavor states are required. In general, whenever we have vanishing entanglement entropy for the initial states, the final state Tsallis and R\'enyi entropies agree at the leading order in the plane wave limit, and is proportional to some elastic scattering cross section/probability.
\begin{table}
\centering
\begin{tabular}{|c|c|c|}
\hline
Subsystem I & The kind of $\sigma$ or $\prob$ & Requiring unentangled initial flavors\\
\hline\hline
$p_\A, f_\A$ & elastic & Yes\\
\hline
$ f_\A, f_\B$ & elastic \& flavor changing & No\\
\hline
$ f_\A$ & elastic \& flavor changing for particle A & Yes\\
\hline
$ f_\B$ & elastic \& flavor changing for particle B & Yes\\
\hline
$ f_\A, p_\B$ & elastic & Yes\\
\hline
$ p_\A$ & elastic & No\\
\hline
$ p_\B$ & elastic & No \\
\hline
\end{tabular}
\caption{The correspondence between the entanglement entropy for different partitions of the final state, and the kind of cross section/probability, as well as whether the initial flavors are required to be not entangled.\label{tab:sum}}
\end{table}

\section{Conclusions}

\label{sec:cao}

We have presented a systematic investigation into 2-to-2 scattering for various bipartition X of the system. When the initial entanglement entropy vanishes, i.e. $\mathcal{E}^{\text{i}}_{n,\text{T}/\text{R},\text{X}} = 0$, the final entanglement entropy in the plane wave limit is
\bea
\mathcal{E}^{\text{f}}_{n,\text{T}/\text{R},\text{X}}  = \frac{n}{n-1} I_0 (|\vec{k}|)\,  \sigma_{\el,\text{Y(X)}} + \ordr (\delta^2_\tv\dtp/|\vec{k}|^3) = \frac{n}{n-1} \prob_{\el,\text{Y(X)}}   + \ordr (\delta^2_\tv\dtp/|\vec{k}|^3),\label{eq:gencon}
\eea
where $\text{Y(X)}$ is the selection of certain flavor configurations for the final state of the elastic scattering, depending on the bipartition X. These are universal relations which do not depend on the underlying dynamics of the QFT that we are considering, and no perturbative expansion in the coupling strength is involved. Neither do the relations depend on the details of the wave packets: at the leading order of the plane wave limit, the information of the wave packets are all factorized out of the computation and represented by a universal overlap integral  $I_0 (|\vec{k}|)$. Furthermore, in Eq. (\ref{eq:ic2g}) we show that $I_0 (|\vec{k}|)$ is the inverse of the area characterizing the transverse size of the wave packets in position space. Then Eq. (\ref{eq:gencon}) first tells us that the entanglement entropy is the total elastic cross section in the unit of the transverse area of the wave packets. This universal relation can thus be interpreted as an area law for the entanglement entropy of a two-body system \cite{Low:2024mrk}.

It has been argued long ago by Froissart and Martin that the total cross section is bounded by $\log^2 s$, where  $\sqrt{s}$ is the CoM energy \cite{Froissart:1961ux,Martin:1962rt}. Cheng and Wu have further shown that for a general theory, in the very high energy limit the cross section grows with energy in a way that saturates the Froissart-Martin bound \cite{Cheng:1969eh,Cheng:1969tje,Cheng:1970bi}, and the contribution of the elastic cross section to the total cross section approaches $1/2$. Such behavior has been observed in experiments of high energy hadron collisions \cite{Block:1984ru,Dremin:2012ke,Pancheri:2016yel,ParticleDataGroup:2022pth}, and through theoretical studies it has been predicted for the future lepton colliders as well \cite{Chen:2016wkt,Han:2020uid,Ruiz:2021tdt}. Our universal relations suggest the entanglement entropy grows with energy in the very high energy limit, which may imply a version of the second law of thermodynamics with respect to collision energy.

This naturally draws comparison with the black hole thermodynamics and the area law of Bekenstein--Hawking \cite{Bekenstein:1973ur,Hawking:1975vcx}. An important realization of the past decade or so is that analogies exist between the scattering of macroscopic objects like black holes, and that of fundamental particles \cite{Holstein:2008sx,Levi:2015msa,Cachazo:2017jef,Guevara:2017csg,Arkani-Hamed:2017jhn,Arkani-Hamed:2019ymq}. Consequently, much effort is being made to apply QFT tools, especially the modern scattering amplitude methods, to computations for gravitational wave experiments \cite{Cheung:2018wkq,Kosower:2018adc,Guevara:2018wpp,Bern:2019nnu,Bern:2019crd,Chung:2019duq,Kalin:2019rwq,Aoude:2020onz,Levi:2020kvb,Levi:2020uwu,Mogull:2020sak,Herrmann:2021tct,Brandhuber:2021eyq,Kim:2021rfj,Edison:2022cdu,Kim:2022pou,Kim:2022bwv,Levi:2022dqm,Levi:2022rrq,Cangemi:2022bew,Edison:2023qvg}. Our area law may provide yet another parallel between black holes and fundamental particles, thus it would be very beneficial to extend our results to incorporate black hole scattering. One possible way to achieve this may be considering the scattering of coherent states \cite{Aoude:2021oqj,Aoude:2023fdm}.

On the other hand, our universal relation in Eq. (\ref{eq:gencon}) can also be seen as a very specific, linear response of the entanglement entropy to the total elastic scattering probability. It would be very interesting to see what happens when we further restrict our selection of final states to specific kinematic configurations, i.e. particles moving in certain directions. Intuitively, the entanglement entropy should then be related to the \textit{differential} cross sections, and it is worth exploring whether any universal relations can be found as well. Furthermore, the fact that different partitions of the system lead to different semi-inclusive cross sections is  fascinating, and one  wonders whether any cross section can be expressed as some entanglement entropy.

Finally, there are many other ways to generalize our study. One may consider other interesting quantities in quantum information theory, e.g. the property of ``magic'' relevant in quantum computing \cite{Robin:2024bdz,White:2024nuc,Brokemeier:2024lhq}. Much can still be explored even if we restrict ourselves to the entanglement entropy. One possibility is to consider inelastic scattering, where the number of outgoing particles may be different from that of incoming particles. There is much to explore about the entanglement beyond bipartite systems \cite{Sakurai:2023nsc}. Yet another possibility is to investigate the case of mixed states, about which we have already derived some interesting results related to unpolarized scattering \cite{Low:2024mrk}. A more thorough exploration may lead to novel predictions that can be tested on colliders.

\begin{acknowledgements}

We thank Rafael Aoude, Tao Han and Nic Pavao for useful discussions. This work  is supported in part by the U.S. Department of Energy, Office of High Energy Physics, under contract DE-AC02-06CH11357 at Argonne, as well as by the U.S. Department of Energy, Office of Nuclear Physics, under grant DE-SC0023522 at Northwestern.

\end{acknowledgements}

\appendix

\section{An additional example for initial wave packets}

\label{app:uwfe}

Here we present a different wave packet configuration compared to the one given in Eqs. (\ref{eq:deltil2}) to (\ref{eq:guxpd}), just to illustrate that our general arguments for the leading order results in the plane wave limit are insensitive to the detailed forms of the wave packets. We will still choose to confine the wave functions in the momentum space strictly inside the cube given in Fig. \ref{fig:wpms}, though now we make the wave functions completely uniform inside the cube. The peak function is given by
\bea
\tilde{\delta}^3 (\vec{p}) = \tilde{\delta}_0  (p_x) \tilde{\delta}_0  (p_y) \tilde{\delta}_0  (p_z),
\eea
where $\tilde{\delta}_0$ is defined in Eq. (\ref{eq:deltil}). The wave function is then
\bea
\psi_{\A/\B} (\vec{p}) = 8(\pi \dtp)^{3/2}  \tilde{\delta}^3 ( \vec{p} - \vec{k}_{\A/\B} ).\label{eq:c1wfn}
\eea
Here we consider head-on collisions, i.e. we set $\vec{b} = \vec{0}$. This is actually the simplest configuration and easiest to compute.

Now we can evaluate the integral $I_0 (|\vec{k}|)$ defined in Eq. (\ref{eq:i0def}). The $p_z$-integration is still given by Eq. (\ref{eq:pzint}), while both $p_x$- and $p_y$-directions give the following:
\bea
\frac{1}{(2 \dtp)^4}\int_{-\dtp}^{\dtp} d (p_1)_{i} \int_{-\dtp}^{\dtp} d (p_2)_{i} \int_{-\dtp}^{\dtp} d (q_1)_{i} \int_{-\dtp}^{\dtp} d (q_2)_{i}\ \delta \left[( \vec{p}_1+ \vec{p}_2- \vec{q}_1 - \vec{q}_2)_{i} \right] = \frac{1}{3\dtp}\ ,
\eea
where $i=x,y$. Including the normalization factor of  $\mathcal{N} = 8 (\pi \dtp)^{3/2}$ by comparing Eqs. (\ref{eq:gwpn}) and (\ref{eq:c1wfn}), we arrive at
\bea
I_0 (|\vec{k}|) =    \frac{|\vec{k}| \sqrt{s}|\mathcal{N}|^4}{(2 \pi)^8  E_{k_\A}  E_{k_{B}}} \frac{E_{k_\A} E_{k_\B}}{ |\vec{k}| \sqrt{s} } \frac{1}{(2 \dtp)^2 (3 \dtp)^2} [1 + \mathcal{O} (\dtp/|\vec{k}|) ]  =  \dtp^2 \left[\frac{4}{9\pi^2} + \mathcal{O} (\dtp/|\vec{k}|) \right],
\eea
which again confirms the scaling given in Eq. (\ref{eq:i0gsc}), as $\delta_\tv = \dtp$ in this configuration.

\section{More examples of diagrams for kinematic data}

\label{app:dkd}

\begin{figure}[tbp]
\centering
\includegraphics[width=0.9\textwidth]{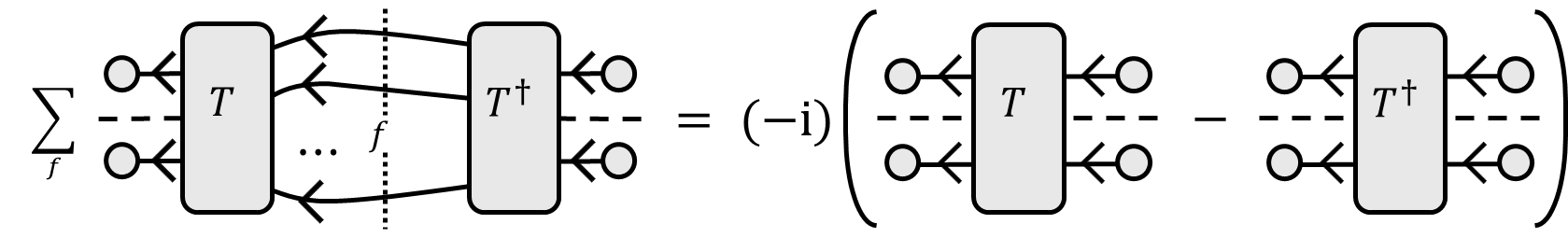}
\caption{The optical theorem for a two body scattering with external wave packets. Notice that the sum over states $\{f \}$ on the left hand side is not restricted to one type-A and one type-B particle, but applies for all possible states, as opposed to all other occasions in our diagrams.\label{fig:optt}}
\end{figure}

In Sec. \ref{sec:fses}, we presented the total scattering probability $\prob_{\text{tot}}$ in Eqs. (\ref{eq:peltot}) and (\ref{eq:ptotitg}), for which we can apply the optical theorem:
\bea
\prob_{\text{tot}} = \< \text{in}| T^\dagger   T |\text{in} \> =2 \< \text{in}| \text{Im}\, T  |\text{in} \>,\label{eq:appopt}
\eea
so that it is converted to tree diagrams with $n_T = 1$, such as Fig. \ref{fig:ont1}. More specifically, the above can be shown diagrammatically in Fig. \ref{fig:optt}. On the other hand, the elastic scattering probability in Eq. (\ref{eq:peltot}) corresponds to the loop diagram in Fig. \ref{fig:cnt2}, with $n_T = 2$, as discussed in Sec. \ref{sec:efm}. As we already see in Sec. \ref{sec:fses}, clearly they should share the same scaling behavior in the $\dtp \to 0$ limit, given by Eq. (\ref{eq:i0gsc}).

\begin{figure}[tbp]%
    \centering
    \subfloat[\centering \label{fig:rhofa231}]{{\includegraphics[width=0.4\textwidth]{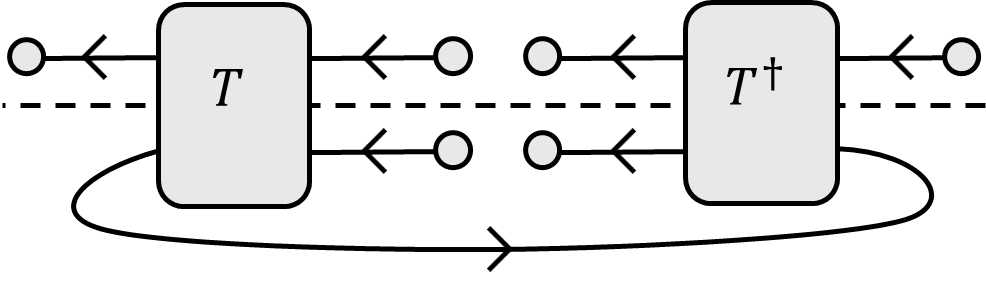} }}%
    \qquad
    \subfloat[\centering \label{fig:rhofa232}]{{\includegraphics[width=0.4\textwidth]{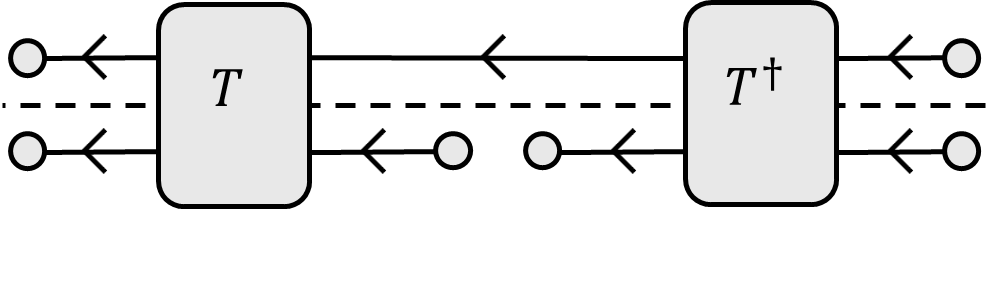} }}%
    \caption{The two terms in the 3rd line of Eq. (\ref{eq:rhoa2c}).}%
\end{figure}

In Sec. \ref{sec:eetp} we have applied our diagrammatic representations to $\tr (\rho_\A)^n$. 
As an example, here we present the complete list of terms for $\tr (\rho_{\A}^{\text{f}})^2$:
\bea
\tr (\rho_{\A}^{\text{f}})^2 &=& \frac{1}{(1 - \prob_{\inel})^2} \Big( 1\non\\
&& -4\ \text{Im} \ \tr_{\text{A}} \left[  \tr_\B \left(T|\text{in} \> \< \text{in} |  \right)  \tr_\B \left(|\text{in} \> \< \text{in} |  \right) \right]  - 2 \ \text{Re} \ \tr_{\text{A}} \left[  \tr_\B \left(T|\text{in} \> \< \text{in} |  \right)   \right]^2\non\\
&&+ 2 \ \tr_\A \left[  \tr_\B \left(T|\text{in} \> \< \text{in} | T^\dagger  \right)  \tr_\B \left(|\text{in} \> \< \text{in} |  \right) \right] + 2\ \tr_{\text{A}} \left[  \tr_\B \left(T|\text{in} \> \< \text{in} |  \right)  \tr_\B \left(|\text{in} \> \< \text{in} | T^\dagger \right) \right]\non\\
&&-4\ \text{Im} \ \tr_{\text{A}} \left[  \tr_\B \left(T|\text{in} \> \< \text{in} |  \right)  \tr_\B \left(T|\text{in} \> \< \text{in} |T^\dagger  \right) \right]  + \tr_{\text{A}} \left[  \tr_\B \left(T|\text{in} \> \< \text{in} | T^\dagger  \right) \right]^2 \Big).\label{eq:rhoa2c}
\eea
The 1st term in the 2nd line of the above scales as $\delta_\tv^2$, and can be represented by $n_T = 1$ tree diagrams such as Fig. \ref{fig:ont1}; the 2nd term in the 2nd line corresponds to the square of Fig. \ref{fig:ont1} and thus scales as $\delta_\tv^4$. The two terms in the 3rd line also scale as $\delta_\tv^4$, which correspond to tree diagrams with $n_T = 2$, given by Figs. \ref{fig:rhofa231} and \ref{fig:rhofa232}. The two terms in the last line of Eq. (\ref{eq:rhoa2c}) scale as $\delta_\tv^6$, with the 1st term given by tree diagrams with $n_T =3$, such as Fig. \ref{fig:ont3}, while the 2nd term is given by the $n_T =4$ loop diagram, Fig. \ref{fig:cnt4}, which is the only loop diagram involved in the computation of $\tr (\rho_{\A}^{\text{f}})^2$.

\bibliographystyle{utphys}
\bibliography{references}

\providecommand{\href}[2]{#2}\begingroup\raggedright\begin{thebibliography}{10}

\bibitem{Einstein:1935rr}
A.~Einstein, B.~Podolsky, and N.~Rosen, ``{Can quantum mechanical description
  of physical reality be considered complete?},''
  \href{http://dx.doi.org/10.1103/PhysRev.47.777}{{\em Phys. Rev.} {\bfseries
  47} (1935) 777--780}.

\bibitem{Bell:1964kc}
J.~S. Bell, ``{On the Einstein-Podolsky-Rosen paradox},''
  \href{http://dx.doi.org/10.1103/PhysicsPhysiqueFizika.1.195}{{\em Physics
  Physique Fizika} {\bfseries 1} (1964) 195--200}.

\bibitem{Nielsen:2012yss}
M.~A. Nielsen and I.~L. Chuang,
  \href{http://dx.doi.org/10.1017/cbo9780511976667}{{\em {Quantum Computation
  and Quantum Information}}}.
\newblock Cambridge University Press, 6, 2012.

\bibitem{ATLAS:2023fsd}
{\bfseries ATLAS} Collaboration, G.~Aad {\em et~al.}, ``{Observation of quantum
  entanglement with top quarks at the ATLAS detector},''
  \href{http://dx.doi.org/10.1038/s41586-024-07824-z}{{\em Nature} {\bfseries
  633} no.~8030, (2024) 542--547},
  \href{http://arxiv.org/abs/2311.07288}{{\ttfamily arXiv:2311.07288
  [hep-ex]}}.

\bibitem{CMS:2024pts}
{\bfseries CMS} Collaboration, A.~Hayrapetyan {\em et~al.}, ``{Observation of
  quantum entanglement in top quark pair production in proton-proton collisions
  at $\sqrt{s}$ = 13 TeV},'' \href{http://arxiv.org/abs/2406.03976}{{\ttfamily
  arXiv:2406.03976 [hep-ex]}}.

\bibitem{Dvali:2014ila}
G.~Dvali, C.~Gomez, R.~S. Isermann, D.~L\"ust, and S.~Stieberger, ``{Black hole
  formation and classicalization in ultra-Planckian 2\textrightarrow{}N
  scattering},'' \href{http://dx.doi.org/10.1016/j.nuclphysb.2015.02.004}{{\em
  Nucl. Phys. B} {\bfseries 893} (2015) 187--235},
  \href{http://arxiv.org/abs/1409.7405}{{\ttfamily arXiv:1409.7405 [hep-th]}}.

\bibitem{Seki:2014cgq}
S.~Seki, I.~Y. Park, and S.-J. Sin, ``{Variation of Entanglement Entropy in
  Scattering Process},''
  \href{http://dx.doi.org/10.1016/j.physletb.2015.02.028}{{\em Phys. Lett. B}
  {\bfseries 743} (2015) 147--153},
  \href{http://arxiv.org/abs/1412.7894}{{\ttfamily arXiv:1412.7894 [hep-th]}}.

\bibitem{Peschanski:2016hgk}
R.~Peschanski and S.~Seki, ``{Entanglement Entropy of Scattering Particles},''
  \href{http://dx.doi.org/10.1016/j.physletb.2016.04.063}{{\em Phys. Lett. B}
  {\bfseries 758} (2016) 89--92},
  \href{http://arxiv.org/abs/1602.00720}{{\ttfamily arXiv:1602.00720
  [hep-th]}}.

\bibitem{Cervera-Lierta:2017tdt}
A.~Cervera-Lierta, J.~I. Latorre, J.~Rojo, and L.~Rottoli, ``{Maximal
  Entanglement in High Energy Physics},''
  \href{http://dx.doi.org/10.21468/SciPostPhys.3.5.036}{{\em SciPost Phys.}
  {\bfseries 3} no.~5, (2017) 036},
  \href{http://arxiv.org/abs/1703.02989}{{\ttfamily arXiv:1703.02989
  [hep-th]}}.

\bibitem{Fan:2017mth}
J.~Fan and X.~Li, ``{Relativistic effect of entanglement in fermion-fermion
  scattering},'' \href{http://dx.doi.org/10.1103/PhysRevD.97.016011}{{\em Phys.
  Rev. D} {\bfseries 97} no.~1, (2018) 016011},
  \href{http://arxiv.org/abs/1712.06237}{{\ttfamily arXiv:1712.06237
  [hep-th]}}.

\bibitem{Beane:2018oxh}
S.~R. Beane, D.~B. Kaplan, N.~Klco, and M.~J. Savage, ``{Entanglement
  Suppression and Emergent Symmetries of Strong Interactions},''
  \href{http://dx.doi.org/10.1103/PhysRevLett.122.102001}{{\em Phys. Rev.
  Lett.} {\bfseries 122} no.~10, (2019) 102001},
  \href{http://arxiv.org/abs/1812.03138}{{\ttfamily arXiv:1812.03138
  [nucl-th]}}.

\bibitem{Tomaras:2019sjq}
T.~N. Tomaras and N.~Toumbas, ``{IR dynamics and entanglement entropy},''
  \href{http://dx.doi.org/10.1103/PhysRevD.101.065006}{{\em Phys. Rev. D}
  {\bfseries 101} no.~6, (2020) 065006},
  \href{http://arxiv.org/abs/1910.07847}{{\ttfamily arXiv:1910.07847
  [hep-th]}}.

\bibitem{Dvali:2020wqi}
G.~Dvali, ``{Entropy Bound and Unitarity of Scattering Amplitudes},''
  \href{http://dx.doi.org/10.1007/JHEP03(2021)126}{{\em JHEP} {\bfseries 03}
  (2021) 126}, \href{http://arxiv.org/abs/2003.05546}{{\ttfamily
  arXiv:2003.05546 [hep-th]}}.

\bibitem{Peschanski:2019yah}
R.~Peschanski and S.~Seki, ``{Evaluation of Entanglement Entropy in High Energy
  Elastic Scattering},''
  \href{http://dx.doi.org/10.1103/PhysRevD.100.076012}{{\em Phys. Rev. D}
  {\bfseries 100} no.~7, (2019) 076012},
  \href{http://arxiv.org/abs/1906.09696}{{\ttfamily arXiv:1906.09696
  [hep-th]}}.

\bibitem{Aoude:2020mlg}
R.~Aoude, M.-Z. Chung, Y.-t. Huang, C.~S. Machado, and M.-K. Tam, ``{Silence of
  Binary Kerr Black Holes},''
  \href{http://dx.doi.org/10.1103/PhysRevLett.125.181602}{{\em Phys. Rev.
  Lett.} {\bfseries 125} no.~18, (2020) 181602},
  \href{http://arxiv.org/abs/2007.09486}{{\ttfamily arXiv:2007.09486
  [hep-th]}}.

\bibitem{Low:2021ufv}
I.~Low and T.~Mehen, ``{Symmetry from entanglement suppression},''
  \href{http://dx.doi.org/10.1103/PhysRevD.104.074014}{{\em Phys. Rev. D}
  {\bfseries 104} no.~7, (2021) 074014},
  \href{http://arxiv.org/abs/2104.10835}{{\ttfamily arXiv:2104.10835
  [hep-th]}}.

\bibitem{Dvali:2021ooc}
G.~Dvali and R.~Venugopalan, ``{Classicalization and unitarization of wee
  partons in QCD and gravity: The CGC-black hole correspondence},''
  \href{http://dx.doi.org/10.1103/PhysRevD.105.056026}{{\em Phys. Rev. D}
  {\bfseries 105} no.~5, (2022) 056026},
  \href{http://arxiv.org/abs/2106.11989}{{\ttfamily arXiv:2106.11989
  [hep-th]}}.

\bibitem{Dvali:2021rlf}
G.~Dvali and O.~Sakhelashvili, ``{Black-hole-like saturons in Gross-Neveu},''
  \href{http://dx.doi.org/10.1103/PhysRevD.105.065014}{{\em Phys. Rev. D}
  {\bfseries 105} no.~6, (2022) 065014},
  \href{http://arxiv.org/abs/2111.03620}{{\ttfamily arXiv:2111.03620
  [hep-th]}}.

\bibitem{Muller:2022htn}
B.~M\"uller and A.~Sch\"afer, ``{Quark-Hadron Transition and Entanglement},''
  \href{http://arxiv.org/abs/2211.16265}{{\ttfamily arXiv:2211.16265
  [hep-ph]}}.

\bibitem{Fedida:2022izl}
S.~Fedida and A.~Serafini, ``{Tree-level entanglement in quantum
  electrodynamics},'' \href{http://dx.doi.org/10.1103/PhysRevD.107.116007}{{\em
  Phys. Rev. D} {\bfseries 107} no.~11, (2023) 116007},
  \href{http://arxiv.org/abs/2209.01405}{{\ttfamily arXiv:2209.01405
  [quant-ph]}}.

\bibitem{Liu:2022grf}
Q.~Liu, I.~Low, and T.~Mehen, ``{Minimal entanglement and emergent symmetries
  in low-energy QCD},''
  \href{http://dx.doi.org/10.1103/PhysRevC.107.025204}{{\em Phys. Rev. C}
  {\bfseries 107} no.~2, (2023) 025204},
  \href{http://arxiv.org/abs/2210.12085}{{\ttfamily arXiv:2210.12085
  [quant-ph]}}.

\bibitem{Cheung:2023hkq}
C.~Cheung, T.~He, and A.~Sivaramakrishnan, ``{Entropy growth in perturbative
  scattering},'' \href{http://dx.doi.org/10.1103/PhysRevD.108.045013}{{\em
  Phys. Rev. D} {\bfseries 108} no.~4, (2023) 045013},
  \href{http://arxiv.org/abs/2304.13052}{{\ttfamily arXiv:2304.13052
  [hep-th]}}.

\bibitem{Hentschinski:2023izh}
M.~Hentschinski, D.~E. Kharzeev, K.~Kutak, and Z.~Tu, ``{Probing the Onset of
  Maximal Entanglement inside the Proton in Diffractive Deep Inelastic
  Scattering},'' \href{http://dx.doi.org/10.1103/PhysRevLett.131.241901}{{\em
  Phys. Rev. Lett.} {\bfseries 131} no.~24, (2023) 241901},
  \href{http://arxiv.org/abs/2305.03069}{{\ttfamily arXiv:2305.03069
  [hep-ph]}}.

\bibitem{Morales:2023gow}
R.~A. Morales, ``{Exploring Bell inequalities and quantum entanglement in
  vector boson scattering},''
  \href{http://dx.doi.org/10.1140/epjp/s13360-023-04784-7}{{\em Eur. Phys. J.
  Plus} {\bfseries 138} no.~12, (2023) 1157},
  \href{http://arxiv.org/abs/2306.17247}{{\ttfamily arXiv:2306.17247
  [hep-ph]}}.

\bibitem{Carena:2023vjc}
M.~Carena, I.~Low, C.~E.~M. Wagner, and M.-L. Xiao, ``{Entanglement
  suppression, enhanced symmetry, and a standard-model-like Higgs boson},''
  \href{http://dx.doi.org/10.1103/PhysRevD.109.L051901}{{\em Phys. Rev. D}
  {\bfseries 109} no.~5, (2024) L051901},
  \href{http://arxiv.org/abs/2307.08112}{{\ttfamily arXiv:2307.08112
  [hep-ph]}}.

\bibitem{Sakurai:2023nsc}
K.~Sakurai and M.~Spannowsky, ``{Three-Body Entanglement in Particle Decays},''
  \href{http://dx.doi.org/10.1103/PhysRevLett.132.151602}{{\em Phys. Rev.
  Lett.} {\bfseries 132} no.~15, (2024) 151602},
  \href{http://arxiv.org/abs/2310.01477}{{\ttfamily arXiv:2310.01477
  [quant-ph]}}.

\bibitem{Liu:2023bnr}
Q.~Liu and I.~Low, ``{Hints of entanglement suppression in hyperon-nucleon
  scattering},'' \href{http://dx.doi.org/10.1016/j.physletb.2024.138899}{{\em
  Phys. Lett. B} {\bfseries 856} (2024) 138899},
  \href{http://arxiv.org/abs/2312.02289}{{\ttfamily arXiv:2312.02289
  [hep-ph]}}.

\bibitem{Hu:2024hex}
T.-R. Hu, S.~Chen, and F.-K. Guo, ``{Entanglement suppression and low-energy
  scattering of heavy mesons},''
  \href{http://dx.doi.org/10.1103/PhysRevD.110.014001}{{\em Phys. Rev. D}
  {\bfseries 110} no.~1, (2024) 014001},
  \href{http://arxiv.org/abs/2404.05958}{{\ttfamily arXiv:2404.05958
  [hep-ph]}}.

\bibitem{Aoude:2024xpx}
R.~Aoude, G.~Elor, G.~N. Remmen, and O.~Sumensari, ``{Positivity in Amplitudes
  from Quantum Entanglement},''
  \href{http://arxiv.org/abs/2402.16956}{{\ttfamily arXiv:2402.16956
  [hep-th]}}.

\bibitem{Kowalska:2024kbs}
K.~Kowalska and E.~M. Sessolo, ``{Entanglement in flavored scalar
  scattering},'' \href{http://dx.doi.org/10.1007/JHEP07(2024)156}{{\em JHEP}
  {\bfseries 07} (2024) 156}, \href{http://arxiv.org/abs/2404.13743}{{\ttfamily
  arXiv:2404.13743 [hep-ph]}}.

\bibitem{Colas:2024ysu}
T.~Colas, J.~Grain, G.~Kaplanek, and V.~Vennin, ``{In-in formalism for the
  entropy of quantum fields in curved spacetimes},''
  \href{http://dx.doi.org/10.1088/1475-7516/2024/08/047}{{\em JCAP} {\bfseries
  08} (2024) 047}, \href{http://arxiv.org/abs/2406.17856}{{\ttfamily
  arXiv:2406.17856 [hep-th]}}.

\bibitem{Low:2024mrk}
I.~Low and Z.~Yin, ``{An Area Law for Entanglement Entropy in Particle
  Scattering},'' \href{http://arxiv.org/abs/2405.08056}{{\ttfamily
  arXiv:2405.08056 [hep-th]}}.

\bibitem{Peskin:1995ev}
M.~E. Peskin and D.~V. Schroeder,
  \href{http://dx.doi.org/10.1201/9780429503559}{{\em {An Introduction to
  quantum field theory}}}.
\newblock Addison-Wesley, Reading, USA, 1995.

\bibitem{Weinberg:1995mt}
S.~Weinberg, \href{http://dx.doi.org/10.1017/CBO9781139644167}{{\em {The
  Quantum theory of fields. Vol. 1: Foundations}}}.
\newblock Cambridge University Press, 6, 2005.

\bibitem{Tsallis}
C.~Tsallis, ``Possible generalization of boltzmann-gibbs statistics,''
  \href{http://dx.doi.org/10.1007/BF01016429}{{\em Journal of Statistical
  Physics} {\bfseries 52} no.~1, (1988) 479--487}.
  \url{https://doi.org/10.1007/BF01016429}.

\bibitem{renyi}
A.~Rényi, {\em {On Measures of Entropy and Information}}.
\newblock in Proceedings of the Fourth Berkeley Symposium on Mathematical
  Statistics and Probability, pp. 547-562, 1960.

\bibitem{Bekenstein:1973ur}
J.~D. Bekenstein, ``{Black holes and entropy},''
  \href{http://dx.doi.org/10.1103/PhysRevD.7.2333}{{\em Phys. Rev. D}
  {\bfseries 7} (1973) 2333--2346}.

\bibitem{Hawking:1975vcx}
S.~W. Hawking, ``{Particle Creation by Black Holes},''
  \href{http://dx.doi.org/10.1007/BF02345020}{{\em Commun. Math. Phys.}
  {\bfseries 43} (1975) 199--220}. [Erratum: Commun.Math.Phys. 46, 206 (1976)].

\bibitem{Srednicki:1993im}
M.~Srednicki, ``{Entropy and area},''
  \href{http://dx.doi.org/10.1103/PhysRevLett.71.666}{{\em Phys. Rev. Lett.}
  {\bfseries 71} (1993) 666--669},
  \href{http://arxiv.org/abs/hep-th/9303048}{{\ttfamily arXiv:hep-th/9303048}}.

\bibitem{Maldacena:1997re}
J.~M. Maldacena, ``{The Large N limit of superconformal field theories and
  supergravity},'' \href{http://dx.doi.org/10.4310/ATMP.1998.v2.n2.a1}{{\em
  Adv. Theor. Math. Phys.} {\bfseries 2} (1998) 231--252},
  \href{http://arxiv.org/abs/hep-th/9711200}{{\ttfamily arXiv:hep-th/9711200}}.

\bibitem{Ryu:2006bv}
S.~Ryu and T.~Takayanagi, ``{Holographic derivation of entanglement entropy
  from AdS/CFT},'' \href{http://dx.doi.org/10.1103/PhysRevLett.96.181602}{{\em
  Phys. Rev. Lett.} {\bfseries 96} (2006) 181602},
  \href{http://arxiv.org/abs/hep-th/0603001}{{\ttfamily arXiv:hep-th/0603001}}.

\bibitem{Eisert:2008ur}
J.~Eisert, M.~Cramer, and M.~B. Plenio, ``{Area laws for the entanglement
  entropy - a review},''
  \href{http://dx.doi.org/10.1103/RevModPhys.82.277}{{\em Rev. Mod. Phys.}
  {\bfseries 82} (2010) 277--306},
  \href{http://arxiv.org/abs/0808.3773}{{\ttfamily arXiv:0808.3773
  [quant-ph]}}.

\bibitem{Dvali:2017nis}
G.~Dvali, ``{Area law microstate entropy from criticality and spherical
  symmetry},'' \href{http://dx.doi.org/10.1103/PhysRevD.97.105005}{{\em Phys.
  Rev. D} {\bfseries 97} no.~10, (2018) 105005},
  \href{http://arxiv.org/abs/1712.02233}{{\ttfamily arXiv:1712.02233
  [hep-th]}}.

\bibitem{Almheiri:2020cfm}
A.~Almheiri, T.~Hartman, J.~Maldacena, E.~Shaghoulian, and A.~Tajdini, ``{The
  entropy of Hawking radiation},''
  \href{http://dx.doi.org/10.1103/RevModPhys.93.035002}{{\em Rev. Mod. Phys.}
  {\bfseries 93} no.~3, (2021) 035002},
  \href{http://arxiv.org/abs/2006.06872}{{\ttfamily arXiv:2006.06872
  [hep-th]}}.

\bibitem{luders1950zustandsanderung}
G.~L{\"u}ders, ``{\"U}ber die zustands{\"a}nderung durch den
  me{\ss}proze{\ss},'' {\em Annalen der Physik} {\bfseries 443} no.~5-8, (1950)
  322--328.

\bibitem{Froissart:1961ux}
M.~Froissart, ``{Asymptotic behavior and subtractions in the Mandelstam
  representation},'' \href{http://dx.doi.org/10.1103/PhysRev.123.1053}{{\em
  Phys. Rev.} {\bfseries 123} (1961) 1053--1057}.

\bibitem{Martin:1962rt}
A.~Martin, ``{Unitarity and high-energy behavior of scattering amplitudes},''
  \href{http://dx.doi.org/10.1103/PhysRev.129.1432}{{\em Phys. Rev.} {\bfseries
  129} (1963) 1432--1436}.

\bibitem{Cheng:1969eh}
H.~Cheng and T.~T. Wu, ``{High-energy elastic scattering in quantum
  electrodynamics},'' \href{http://dx.doi.org/10.1103/PhysRevLett.22.666}{{\em
  Phys. Rev. Lett.} {\bfseries 22} (1969) 666}.

\bibitem{Cheng:1969tje}
H.~Cheng and T.~T. Wu, ``{Impact factor and exponentiation in high-energy
  scattering processes},''
  \href{http://dx.doi.org/10.1103/PhysRev.186.1611}{{\em Phys. Rev.} {\bfseries
  186} (1969) 1611--1618}.

\bibitem{Cheng:1970bi}
H.~Cheng and T.~T. Wu, ``{Limit of Cross-Sections at Infinite Energy},''
  \href{http://dx.doi.org/10.1103/PhysRevLett.24.1456}{{\em Phys. Rev. Lett.}
  {\bfseries 24} (1970) 1456--1460}.

\bibitem{Block:1984ru}
M.~M. Block and R.~N. Cahn, ``{High-Energy $p \bar{p}$ and $p p$ Forward
  Elastic Scattering and Total Cross-Sections},''
  \href{http://dx.doi.org/10.1103/RevModPhys.57.563}{{\em Rev. Mod. Phys.}
  {\bfseries 57} (1985) 563}.

\bibitem{Dremin:2012ke}
I.~M. Dremin, ``{Elastic scattering of hadrons},''
  \href{http://dx.doi.org/10.3367/UFNe.0183.201301a.0003}{{\em Phys. Usp.}
  {\bfseries 56} (2013) 3--28},
  \href{http://arxiv.org/abs/1206.5474}{{\ttfamily arXiv:1206.5474 [hep-ph]}}.

\bibitem{Pancheri:2016yel}
G.~Pancheri and Y.~N. Srivastava, ``{Introduction to the physics of the total
  cross-section at LHC}: {A Review of Data and Models},''
  \href{http://dx.doi.org/10.1140/epjc/s10052-016-4585-8}{{\em Eur. Phys. J. C}
  {\bfseries 77} no.~3, (2017) 150},
  \href{http://arxiv.org/abs/1610.10038}{{\ttfamily arXiv:1610.10038
  [hep-ph]}}.

\bibitem{ParticleDataGroup:2022pth}
{\bfseries Particle Data Group} Collaboration, R.~L. Workman {\em et~al.},
  ``{Review of Particle Physics},''
  \href{http://dx.doi.org/10.1093/ptep/ptac097}{{\em PTEP} {\bfseries 2022}
  (2022) 083C01}.

\bibitem{Chen:2016wkt}
J.~Chen, T.~Han, and B.~Tweedie, ``{Electroweak Splitting Functions and High
  Energy Showering},'' \href{http://dx.doi.org/10.1007/JHEP11(2017)093}{{\em
  JHEP} {\bfseries 11} (2017) 093},
  \href{http://arxiv.org/abs/1611.00788}{{\ttfamily arXiv:1611.00788
  [hep-ph]}}.

\bibitem{Han:2020uid}
T.~Han, Y.~Ma, and K.~Xie, ``{High energy leptonic collisions and electroweak
  parton distribution functions},''
  \href{http://dx.doi.org/10.1103/PhysRevD.103.L031301}{{\em Phys. Rev. D}
  {\bfseries 103} no.~3, (2021) L031301},
  \href{http://arxiv.org/abs/2007.14300}{{\ttfamily arXiv:2007.14300
  [hep-ph]}}.

\bibitem{Ruiz:2021tdt}
R.~Ruiz, A.~Costantini, F.~Maltoni, and O.~Mattelaer, ``{The Effective Vector
  Boson Approximation in high-energy muon collisions},''
  \href{http://dx.doi.org/10.1007/JHEP06(2022)114}{{\em JHEP} {\bfseries 06}
  (2022) 114}, \href{http://arxiv.org/abs/2111.02442}{{\ttfamily
  arXiv:2111.02442 [hep-ph]}}.

\bibitem{Holstein:2008sx}
B.~R. Holstein and A.~Ross, ``{Spin Effects in Long Range Gravitational
  Scattering},'' \href{http://arxiv.org/abs/0802.0716}{{\ttfamily
  arXiv:0802.0716 [hep-ph]}}.

\bibitem{Levi:2015msa}
M.~Levi and J.~Steinhoff, ``{Spinning gravitating objects in the effective
  field theory in the post-Newtonian scheme},''
  \href{http://dx.doi.org/10.1007/JHEP09(2015)219}{{\em JHEP} {\bfseries 09}
  (2015) 219}, \href{http://arxiv.org/abs/1501.04956}{{\ttfamily
  arXiv:1501.04956 [gr-qc]}}.

\bibitem{Cachazo:2017jef}
F.~Cachazo and A.~Guevara, ``{Leading Singularities and Classical Gravitational
  Scattering},'' \href{http://dx.doi.org/10.1007/JHEP02(2020)181}{{\em JHEP}
  {\bfseries 02} (2020) 181}, \href{http://arxiv.org/abs/1705.10262}{{\ttfamily
  arXiv:1705.10262 [hep-th]}}.

\bibitem{Guevara:2017csg}
A.~Guevara, ``{Holomorphic Classical Limit for Spin Effects in Gravitational
  and Electromagnetic Scattering},''
  \href{http://dx.doi.org/10.1007/JHEP04(2019)033}{{\em JHEP} {\bfseries 04}
  (2019) 033}, \href{http://arxiv.org/abs/1706.02314}{{\ttfamily
  arXiv:1706.02314 [hep-th]}}.

\bibitem{Arkani-Hamed:2017jhn}
N.~Arkani-Hamed, T.-C. Huang, and Y.-t. Huang, ``{Scattering amplitudes for all
  masses and spins},'' \href{http://dx.doi.org/10.1007/JHEP11(2021)070}{{\em
  JHEP} {\bfseries 11} (2021) 070},
  \href{http://arxiv.org/abs/1709.04891}{{\ttfamily arXiv:1709.04891
  [hep-th]}}.

\bibitem{Arkani-Hamed:2019ymq}
N.~Arkani-Hamed, Y.-t. Huang, and D.~O'Connell, ``{Kerr black holes as
  elementary particles},''
  \href{http://dx.doi.org/10.1007/JHEP01(2020)046}{{\em JHEP} {\bfseries 01}
  (2020) 046}, \href{http://arxiv.org/abs/1906.10100}{{\ttfamily
  arXiv:1906.10100 [hep-th]}}.

\bibitem{Cheung:2018wkq}
C.~Cheung, I.~Z. Rothstein, and M.~P. Solon, ``{From Scattering Amplitudes to
  Classical Potentials in the Post-Minkowskian Expansion},''
  \href{http://dx.doi.org/10.1103/PhysRevLett.121.251101}{{\em Phys. Rev.
  Lett.} {\bfseries 121} no.~25, (2018) 251101},
  \href{http://arxiv.org/abs/1808.02489}{{\ttfamily arXiv:1808.02489
  [hep-th]}}.

\bibitem{Kosower:2018adc}
D.~A. Kosower, B.~Maybee, and D.~O'Connell, ``{Amplitudes, Observables, and
  Classical Scattering},''
  \href{http://dx.doi.org/10.1007/JHEP02(2019)137}{{\em JHEP} {\bfseries 02}
  (2019) 137}, \href{http://arxiv.org/abs/1811.10950}{{\ttfamily
  arXiv:1811.10950 [hep-th]}}.

\bibitem{Guevara:2018wpp}
A.~Guevara, A.~Ochirov, and J.~Vines, ``{Scattering of Spinning Black Holes
  from Exponentiated Soft Factors},''
  \href{http://dx.doi.org/10.1007/JHEP09(2019)056}{{\em JHEP} {\bfseries 09}
  (2019) 056}, \href{http://arxiv.org/abs/1812.06895}{{\ttfamily
  arXiv:1812.06895 [hep-th]}}.

\bibitem{Bern:2019nnu}
Z.~Bern, C.~Cheung, R.~Roiban, C.-H. Shen, M.~P. Solon, and M.~Zeng,
  ``{Scattering Amplitudes and the Conservative Hamiltonian for Binary Systems
  at Third Post-Minkowskian Order},''
  \href{http://dx.doi.org/10.1103/PhysRevLett.122.201603}{{\em Phys. Rev.
  Lett.} {\bfseries 122} no.~20, (2019) 201603},
  \href{http://arxiv.org/abs/1901.04424}{{\ttfamily arXiv:1901.04424
  [hep-th]}}.

\bibitem{Bern:2019crd}
Z.~Bern, C.~Cheung, R.~Roiban, C.-H. Shen, M.~P. Solon, and M.~Zeng, ``{Black
  Hole Binary Dynamics from the Double Copy and Effective Theory},''
  \href{http://dx.doi.org/10.1007/JHEP10(2019)206}{{\em JHEP} {\bfseries 10}
  (2019) 206}, \href{http://arxiv.org/abs/1908.01493}{{\ttfamily
  arXiv:1908.01493 [hep-th]}}.

\bibitem{Chung:2019duq}
M.-Z. Chung, Y.-T. Huang, and J.-W. Kim, ``{Classical potential for general
  spinning bodies},'' \href{http://dx.doi.org/10.1007/JHEP09(2020)074}{{\em
  JHEP} {\bfseries 09} (2020) 074},
  \href{http://arxiv.org/abs/1908.08463}{{\ttfamily arXiv:1908.08463
  [hep-th]}}.

\bibitem{Kalin:2019rwq}
G.~K\"alin and R.~A. Porto, ``{From Boundary Data to Bound States},''
  \href{http://dx.doi.org/10.1007/JHEP01(2020)072}{{\em JHEP} {\bfseries 01}
  (2020) 072}, \href{http://arxiv.org/abs/1910.03008}{{\ttfamily
  arXiv:1910.03008 [hep-th]}}.

\bibitem{Aoude:2020onz}
R.~Aoude, K.~Haddad, and A.~Helset, ``{On-shell heavy particle effective
  theories},'' \href{http://dx.doi.org/10.1007/JHEP05(2020)051}{{\em JHEP}
  {\bfseries 05} (2020) 051}, \href{http://arxiv.org/abs/2001.09164}{{\ttfamily
  arXiv:2001.09164 [hep-th]}}.

\bibitem{Levi:2020kvb}
M.~Levi, A.~J. Mcleod, and M.~Von~Hippel, ``{N$^{3}$LO gravitational spin-orbit
  coupling at order G$^{4}$},''
  \href{http://dx.doi.org/10.1007/JHEP07(2021)115}{{\em JHEP} {\bfseries 07}
  (2021) 115}, \href{http://arxiv.org/abs/2003.02827}{{\ttfamily
  arXiv:2003.02827 [hep-th]}}.

\bibitem{Levi:2020uwu}
M.~Levi, A.~J. Mcleod, and M.~Von~Hippel, ``{N$^{3}$LO gravitational
  quadratic-in-spin interactions at G$^{4}$},''
  \href{http://dx.doi.org/10.1007/JHEP07(2021)116}{{\em JHEP} {\bfseries 07}
  (2021) 116}, \href{http://arxiv.org/abs/2003.07890}{{\ttfamily
  arXiv:2003.07890 [hep-th]}}.

\bibitem{Mogull:2020sak}
G.~Mogull, J.~Plefka, and J.~Steinhoff, ``{Classical black hole scattering from
  a worldline quantum field theory},''
  \href{http://dx.doi.org/10.1007/JHEP02(2021)048}{{\em JHEP} {\bfseries 02}
  (2021) 048}, \href{http://arxiv.org/abs/2010.02865}{{\ttfamily
  arXiv:2010.02865 [hep-th]}}.

\bibitem{Herrmann:2021tct}
E.~Herrmann, J.~Parra-Martinez, M.~S. Ruf, and M.~Zeng, ``{Radiative classical
  gravitational observables at $ \mathcal{O} $(G$^{3}$) from scattering
  amplitudes},'' \href{http://dx.doi.org/10.1007/JHEP10(2021)148}{{\em JHEP}
  {\bfseries 10} (2021) 148}, \href{http://arxiv.org/abs/2104.03957}{{\ttfamily
  arXiv:2104.03957 [hep-th]}}.

\bibitem{Brandhuber:2021eyq}
A.~Brandhuber, G.~Chen, G.~Travaglini, and C.~Wen, ``{Classical gravitational
  scattering from a gauge-invariant double copy},''
  \href{http://dx.doi.org/10.1007/JHEP10(2021)118}{{\em JHEP} {\bfseries 10}
  (2021) 118}, \href{http://arxiv.org/abs/2108.04216}{{\ttfamily
  arXiv:2108.04216 [hep-th]}}.

\bibitem{Kim:2021rfj}
J.-W. Kim, M.~Levi, and Z.~Yin, ``{Quadratic-in-spin interactions at fifth
  post-Newtonian order probe new physics},''
  \href{http://dx.doi.org/10.1016/j.physletb.2022.137410}{{\em Phys. Lett. B}
  {\bfseries 834} (2022) 137410},
  \href{http://arxiv.org/abs/2112.01509}{{\ttfamily arXiv:2112.01509
  [hep-th]}}.

\bibitem{Edison:2022cdu}
A.~Edison and M.~Levi, ``{A tale of tails through generalized unitarity},''
  \href{http://dx.doi.org/10.1016/j.physletb.2022.137634}{{\em Phys. Lett. B}
  {\bfseries 837} (2023) 137634},
  \href{http://arxiv.org/abs/2202.04674}{{\ttfamily arXiv:2202.04674
  [hep-th]}}.

\bibitem{Kim:2022pou}
J.-W. Kim, M.~Levi, and Z.~Yin, ``{N$^{3}$LO spin-orbit interaction via the EFT
  of spinning gravitating objects},''
  \href{http://dx.doi.org/10.1007/JHEP05(2023)184}{{\em JHEP} {\bfseries 05}
  (2023) 184}, \href{http://arxiv.org/abs/2208.14949}{{\ttfamily
  arXiv:2208.14949 [hep-th]}}.

\bibitem{Kim:2022bwv}
J.-W. Kim, M.~Levi, and Z.~Yin, ``{N$^{3}$LO quadratic-in-spin interactions for
  generic compact binaries},''
  \href{http://dx.doi.org/10.1007/JHEP03(2023)098}{{\em JHEP} {\bfseries 03}
  (2023) 098}, \href{http://arxiv.org/abs/2209.09235}{{\ttfamily
  arXiv:2209.09235 [hep-th]}}.

\bibitem{Levi:2022dqm}
M.~Levi, R.~Morales, and Z.~Yin, ``{From the EFT of spinning gravitating
  objects to Poincar\'e and gauge invariance at the 4.5PN precision
  frontier},'' \href{http://dx.doi.org/10.1007/JHEP09(2023)090}{{\em JHEP}
  {\bfseries 09} (2023) 090}, \href{http://arxiv.org/abs/2210.17538}{{\ttfamily
  arXiv:2210.17538 [hep-th]}}.

\bibitem{Levi:2022rrq}
M.~Levi and Z.~Yin, ``{Completing the fifth PN precision frontier via the EFT
  of spinning gravitating objects},''
  \href{http://dx.doi.org/10.1007/JHEP04(2023)079}{{\em JHEP} {\bfseries 04}
  (2023) 079}, \href{http://arxiv.org/abs/2211.14018}{{\ttfamily
  arXiv:2211.14018 [hep-th]}}.

\bibitem{Cangemi:2022bew}
L.~Cangemi, M.~Chiodaroli, H.~Johansson, A.~Ochirov, P.~Pichini, and
  E.~Skvortsov, ``{Kerr Black Holes From Massive Higher-Spin Gauge Symmetry},''
  \href{http://dx.doi.org/10.1103/PhysRevLett.131.221401}{{\em Phys. Rev.
  Lett.} {\bfseries 131} no.~22, (2023) 221401},
  \href{http://arxiv.org/abs/2212.06120}{{\ttfamily arXiv:2212.06120
  [hep-th]}}.

\bibitem{Edison:2023qvg}
A.~Edison and M.~Levi, ``{Higher-order tails and RG flows due to scattering of
  gravitational radiation from binary inspirals},''
  \href{http://dx.doi.org/10.1007/JHEP08(2024)161}{{\em JHEP} {\bfseries 08}
  (2024) 161}, \href{http://arxiv.org/abs/2310.20066}{{\ttfamily
  arXiv:2310.20066 [hep-th]}}.

\bibitem{Aoude:2021oqj}
R.~Aoude and A.~Ochirov, ``{Classical observables from coherent-spin
  amplitudes},'' \href{http://dx.doi.org/10.1007/JHEP10(2021)008}{{\em JHEP}
  {\bfseries 10} (2021) 008}, \href{http://arxiv.org/abs/2108.01649}{{\ttfamily
  arXiv:2108.01649 [hep-th]}}.

\bibitem{Aoude:2023fdm}
R.~Aoude and A.~Ochirov, ``{Gravitational partial-wave absorption from
  scattering amplitudes},''
  \href{http://dx.doi.org/10.1007/JHEP12(2023)103}{{\em JHEP} {\bfseries 12}
  (2023) 103}, \href{http://arxiv.org/abs/2307.07504}{{\ttfamily
  arXiv:2307.07504 [hep-th]}}.

\bibitem{Robin:2024bdz}
C.~E.~P. Robin and M.~J. Savage, ``{The Magic in Nuclear and Hypernuclear
  Forces},'' \href{http://arxiv.org/abs/2405.10268}{{\ttfamily arXiv:2405.10268
  [nucl-th]}}.

\bibitem{White:2024nuc}
C.~D. White and M.~J. White, ``{The magic of entangled top quarks},''
  \href{http://arxiv.org/abs/2406.07321}{{\ttfamily arXiv:2406.07321
  [hep-ph]}}.

\bibitem{Brokemeier:2024lhq}
F.~Br\"okemeier, S.~M. Hengstenberg, J.~W.~T. Keeble, C.~E.~P. Robin, F.~Rocco,
  and M.~J. Savage, ``{Quantum Magic and Multi-Partite Entanglement in the
  Structure of Nuclei},'' \href{http://arxiv.org/abs/2409.12064}{{\ttfamily
  arXiv:2409.12064 [nucl-th]}}.

\end{thebibliography}\endgroup

\end{document}